\documentclass[11pt]{article}
\usepackage{amsmath}
\usepackage{color}
\usepackage{latexsym,amsmath,amssymb}
\usepackage{graphicx,graphics,subfigure}
\renewcommand{\baselinestretch}{1.46}

\newtheorem{lemma}{Lemma}
\newtheorem{example}{Example}

\begin{document}

\title{A Computationally Stable Approach to Gaussian Process Interpolation of Deterministic Computer Simulation Data}

\renewcommand{\baselinestretch}{1.0}
\author{Pritam Ranjan$^1$, Ronald Haynes$^2$ and Richard Karsten$^1$ \vspace{0.5cm}\\
$^1$ \small Department of Mathematics and Statistics,
Acadia University, NS, Canada\\ $^2$ \small Department of Mathematics and Statistics,
Memorial University, NL, Canada\\
\small (pritam.ranjan@acadiau.ca, rhaynes@mun.ca, richard.karsten@acadiau.ca)
}


\date{}

\maketitle
\begin{abstract}
For many expensive deterministic computer simulators, the outputs do not have replication error and the desired metamodel (or statistical emulator) is an interpolator of the observed data. Realizations of Gaussian spatial processes (GP) are commonly used to model such simulator outputs. Fitting a GP model to $n$ data points requires the computation of the inverse and determinant of $n \times n$ correlation matrices, $R$, that are sometimes computationally unstable due to near-singularity of $R$. This happens if any pair of design points are very close together in the input space. The popular approach to overcome near-singularity is to introduce a small nugget (or jitter) parameter in the model that is estimated along with other model parameters. The inclusion of a nugget in the model often causes unnecessary over-smoothing of the data. In this paper, we propose a lower bound on the nugget that minimizes the over-smoothing and an iterative regularization approach to construct a predictor that further improves the interpolation accuracy. We also show that the proposed predictor converges to the GP interpolator.\\


\end{abstract}
\renewcommand{\baselinestretch}{1.4}

\section{Introduction}\label{sec:intro}
Computer simulators are often used to model complex physical and engineering processes that are either too expensive or time consuming to observe. A simulator is said to be deterministic if the replicate runs of the same inputs will yield identical responses. For the last
few decades, deterministic simulators have been widely used to model physical processes. For instance, Kumar and Davidson (1978) used deterministic simulation models for comparing the performance of highly concurrent computers; Su et al. (1996) used generalized linear regression models to design a lamp filament via a deterministic finite-element computer code; Aslett et al. (1998) discuss an optimization problem for a deterministic circuit simulator; several deterministic simulators are being used for analyzing biochemical networks (see Bergmann and Sauro 2008 for references). On the other hand, there are cases where stochastic (non-deterministic) simulators are preferred due to unavoidable biases (e.g., Poole and Raftery 2000). In spite of the recent interest in stochastic simulators, deterministic simulators are still being actively used. For instance, Medina, Moreno and Royo (2005) demonstrate the preference of deterministic traffic simulators over their stochastic counterparts. In this paper, we assume that the simulator under consideration is deterministic up to working precision and the scientist is
confident about the validity of the simulator.

Sacks, Welch, Mitchell and Wynn (1989) proposed modeling (or emulating) such an expensive deterministic simulator as a realization of a Gaussian stochastic process (GP). An emulator of a deterministic simulator is desired to be an interpolator of the observed data (e.g., Sacks et al. 1989; Van Beers and Kleijnen 2004). For the problem that motivated this work, the objective is to emulate the average extractable tidal power as a function of the turbine locations in the Bay of Fundy, Nova Scotia, Canada. The deterministic computer simulator for the tidal power model is a numerical solver of a complex system of partial differential equations, and we accept the simulator as a valid representation of the tidal power.

In this paper, we discuss a computational issue in building the GP based emulator for a deterministic simulator. Fitting a GP model to $n$ data points using either a maximum likelihood technique or a Bayesian approach requires the computation of the determinant and inverse of several $n \times n$ correlation matrices, $R$. Although the correlation matrices are positive definite by definition, near-singularity (also referred to as ill-conditioning) of these matrices is a common problem in fitting GP models. Ababou, Bagtzoglou and Wood (1994) study the relationship of a uniform grid to the ill-conditioning and quality of model fit for various covariance models. Barton and Salagame (1997) study the effect of experimental design on the ill-conditioning of kriging models. Jones, Schonlau and Welch (1998) used the singular value decomposition to overcome the near-singularity of $R$.
Booker (2000) used the sum of independent GPs to overcome near-singularity for multi-stage adaptive designs in kriging models. A more popular solution to overcome near-singularity is to introduce a \emph{nugget or jitter} parameter, $\delta$, in the model (e.g., Sacks et al. 1989; Neal 1997; Booker et al. 1999; Santner, Williams and Notz 2003; Gramacy and Lee 2008) that is estimated along with other model parameters. However, adding a nugget to the model introduces additional smoothing in the predictor and as a result the predictor is no longer an interpolator.

Here, we first propose a lower bound on the nugget ($\delta_{lb}$) that minimizes the additional over-smoothing. Second, an iterative approach is developed to enable the construction of a new predictor that further improves the interpolation as well as the prediction (at unsampled design points) accuracy. We also show that the proposed predictor converges to an interpolator. Although an arbitrary nugget $(0 < \delta < 1)$ can be used in the iterative approach, the rate of convergence (i.e., the number of iterations required to reach certain tolerance) depends on the magnitude of the nugget. To this effect, the proposed lower bound $\delta_{lb}$ significantly reduces the number of iterations required. This feature is particularly desirable for implementation.

The paper is organized as follows. Section~\ref{sec:motivating} presents the tidal power modeling example. In Section~\ref{sec:background}, we review the GP model, a computational issue in fitting the model, and the popular approach to overcome near-singularity. Section~\ref{sec:nugget} presents the new lower bound for the nugget that is required to achieve well-conditioned correlation matrices and minimize unnecessary over-smoothing. In Section~\ref{sec:iterative}, we develop the iterative approach for constructing a more accurate predictor. Several examples are presented in Section~\ref{sec:examples} to illustrate the performance of our proposed predictor over the one obtained using the popular approach. Finally, we conclude the paper with some remarks on the numerical issues and recommendations for practitioners in Section~\ref{sec:discussion}.

\section{Motivating example}\label{sec:motivating}
\begin{figure}[h!]\centering
\subfigure[The upper Bay of Fundy]{\label{fig:BOF1a}
\includegraphics[height=2.6in,width=3.7in]{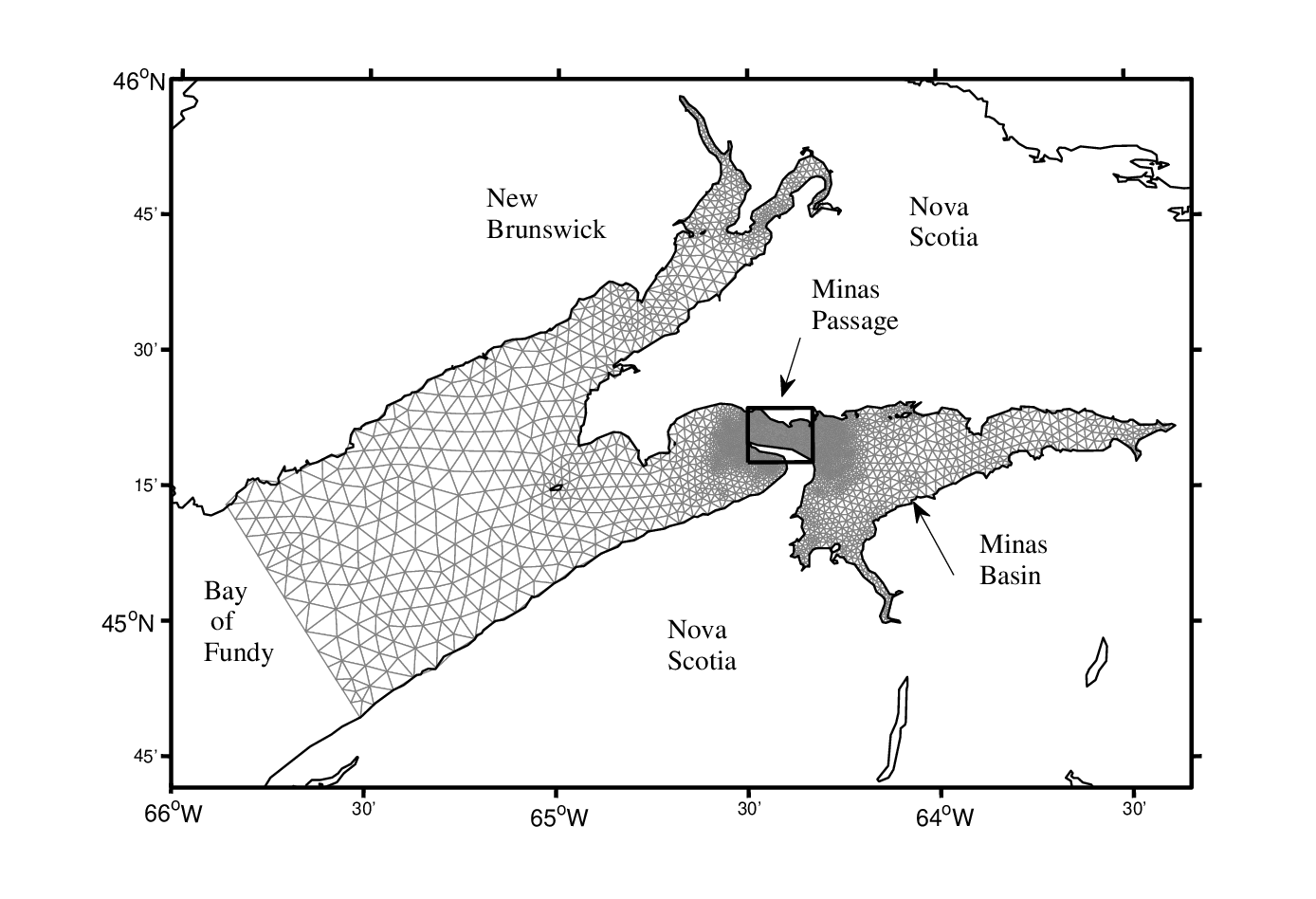} \hspace{-1.2cm}}
\subfigure[The Minas Passage]{\label{fig:BOF1b} \includegraphics[height=2.5in,width=3.0in]{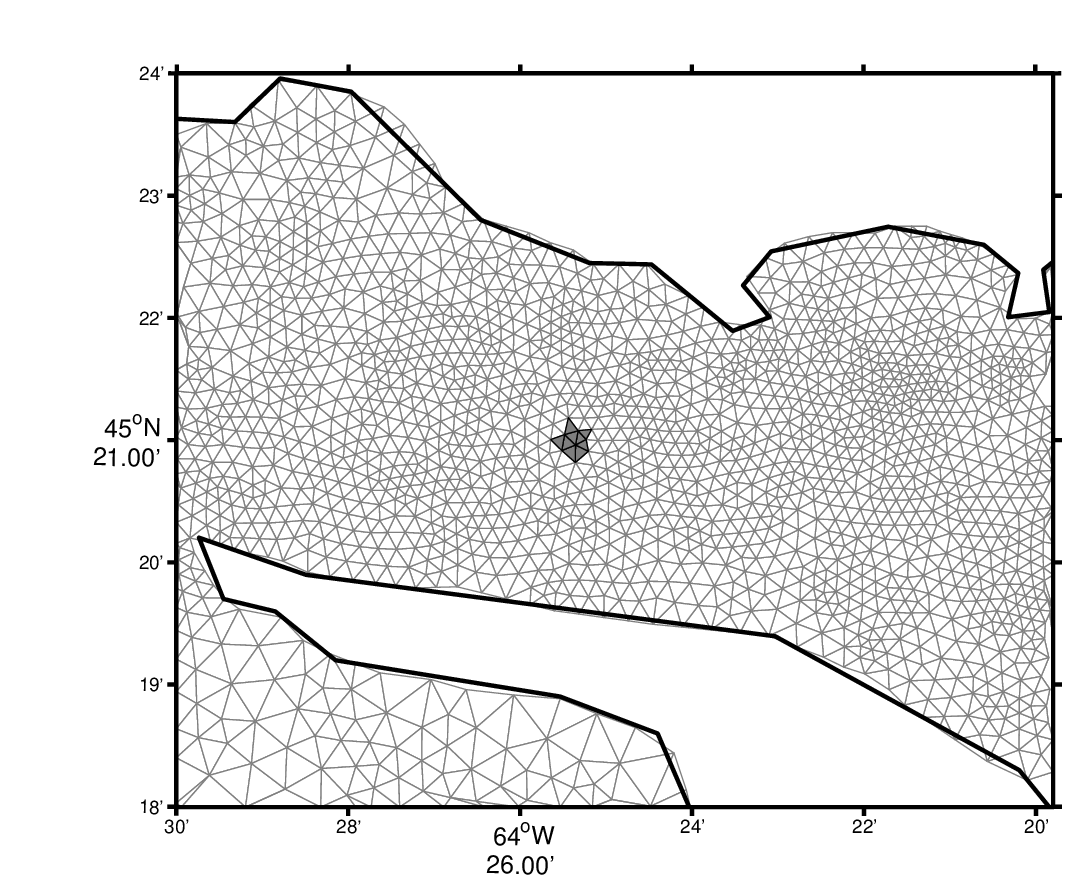}}
\caption{Figure (a) shows the triangular grid used in the FVCOM model for simulating tides in the upper Bay of Fundy. The small box in the center surrounds the Minas Passage shown in (b). The shaded triangles in the center of (b) represent a possible turbine location.}\label{fig:BOF1}
\end{figure}

The Bay of Fundy, located between New Brunswick and Nova Scotia, Canada, with a small portion touching Maine, USA, is world famous for its high tides. In the upper portion of the Bay of Fundy
(see Figure~\ref{fig:BOF1}(a)), the difference in water level between high tide and low tide can be as much as 17 meters. The high tides in this region are a result of a resonance, with the natural period of the Bay of Fundy very close to the period of the principal lunar tide. This results in very regular tides in the Bay of Fundy with a high tide every 12.42 hours. The incredible energy in these tides has meant that the region has significant potential for extracting tidal power (Greenberg 1979; Karsten, McMillan, Lickley and Haynes 2008 (hereafter KMLH)).

Though the notion of harnessing tidal power from the Bay of Fundy is not new, earlier proposed methods of harvesting the much needed green electrical energy involved building a barrage or dam. This method was considered infeasible for a variety of economic and environmental reasons. Recently, there has been rapid technological development of in-steam tidal turbines.
These devices act much like wind turbines, with individual turbines placed in regions of strong tidal currents. Individual turbines can be up to 20 m in diameter and can produce over 1 MW of power. Ideally, these turbines would produce a predictable and renewable source of power with less of an impact on the environment than a dam. KMLH examined the power potential of farms of such turbines across the Minas Passage (Figure~\ref{fig:BOF1}(b)) where the tidal currents are strongest. They found that the potential extractable power is much higher than previous estimates and that the environmental impacts of extracting power can be greatly reduced by extracting only a portion of the maximum power available. The simulations in KMLH did not represent individual turbines and left open the question of how to optimally place turbines. In this paper, we emulate the KMLH numerical model to examine the placement of turbines to maximize the power output.

We numerically simulate the tides as in KMLH by solving the 2D shallow water equations using the Finite-Volume Coastal Ocean Model (FVCOM) with a triangular grid on the upper Bay of Fundy (see Figure \ref{fig:BOF1a}). Since the grid triangles differ in size and orientation, the $i$-th turbine was modeled on the set of all triangular elements whose centers lie within 250 m of $(x_i, y_i)$. A possible turbine location is shown in Figure \ref{fig:BOF1b}. The triangular grid was developed by David Greenberg and colleagues at the Bedford Institute of Oceanography, NS, Canada. The details of FVCOM can be found in Chen et al. (2006).

Using this set up, the estimate of the electric power that can be harnessed through a turbine at a particular location $(x, y)$ over a tidal cycle $T=12.42$ hours is obtained by the simulator in KMLH. The average tidal power at the location $(x, y)$ is given by
$$ \bar{P}(x, y) = \frac{1}{T} \int^T_0 P(t; x, y)dt\, ,$$
where $P(t; x, y)$ is the extractable power output at time $t$ and location $(x, y)$. The process is deterministic up to the machine precision, and the main objective is to emulate $\bar{P}(x, y)$.

It turns out that the GP model fitted to the simulator output at $n=100$ points (chosen using a space-filling design criterion) is not an interpolator and results in an over-smoothed emulator (see Example~4 for details). This is undesirable as the ocean modelers are interested in an emulator that interpolates their simulator. This emulator will be used to obtain estimates of both the maximizer of the power function (i.e., the location where to put the turbine) and the extractable power at this location. The manufacturing and installation cost of the initial prototype turbine is very high (roughly 20 million dollars). Since the over-smoothed emulator can underestimate the maximum extractable power, a good approximation of the attainable power function can be helpful in saving the cost of a few turbines. Example~4 shows that the proposed approach leads to a more accurate estimate of the maximum extractable power.

\section{Background review}\label{sec:background}
\subsection{Gaussian process model}\label{sec:GP_model}
Let the $i$-th input and output of the computer simulator be denoted by a $d$-dimensional vector, $x_i=(x_{i1},...,x_{id})$, and the univariate response, $y_i = y(x_i)$, respectively. The experiment design $D_0 = \{x_1,...,x_n\}$ is the set of $n$ input trials. The outputs of the simulation trials are held in the $n$-dimensional vector $Y=y(D_0)=(y_1, y_2,\dots,y_n)'$. The simulator output, $y(x_i)$, is modeled as
\begin{equation}\label{GASP}
    y(x_i) = \mu + z(x_i);
 \qquad i=1,...,n,
\end{equation}
where $\mu$ is the overall mean, and $z(x_i)$ is a GP with $E(z(x_i))=0$, $Var(z(x_i))= \sigma^2_z$, and Cov$(z(x_i),z(x_j)) = \sigma^2_z R_{ij}$. In general, $y(D_0)$ has a multivariate normal distribution, $N_n({\bf 1_n}\mu, \Sigma)$, where $\Sigma = V(D_0|y(D_0)) = \sigma^2_zR $, and ${\bf 1_n}$ is a $n \times 1$ vector of all ones (see Sacks et al. 1989, and Jones et al. 1998 for details). Although there are several choices for the correlation function, we focus on the Gaussian correlation because of its properties like smoothness (or differentiability in mean square sense) and popularity in other areas like machine learning (radial basis kernels) and geostatistics (kriging). For a detailed discussion on correlation functions see Stein (1999), Santner, Williams and Notz (2003), and Rasmussen and Williams (2006). The Gaussian correlation function is a special case ($p_k=2$ for all $k$) of the power exponential correlation family
\begin{equation}\label{corr}
   R_{ij} = \mbox{corr}(z(x_i),z(x_j)) =
   \prod_{k=1}^{d}\exp \left\{-\theta_k|x_{ik}-x_{jk}|^{p_k}\right\},
   \quad \mbox{ for all }  \quad i,j,
\end{equation}
where $\theta = (\theta_1,...,\theta_d)$ is the vector of hyper-parameters, and $p_k \in (0,2]$ is the smoothness parameter. As discussed in Section~7, the results developed in this paper may vary slightly when other correlation structures are used instead of the Gaussian correlation.

We use the GP model with Gaussian correlation function to predict responses at any unsampled design point $x^*$, however, the theory developed here is also valid for other correlation structures in the power exponential family (see Section~7 for more details). Following the maximum likelihood approach, the best linear unbiased predictor (BLUP) at $x^*$ is
\begin{eqnarray}\label{yhat}
   \hat{y}(x^*) &=& \hat{\mu} + r'R^{-1}(Y-{\bf1_n}\hat{\mu})
     = \left[ \frac{(1-r'R^{-1}\mathbf{1}_n)}{\mathbf{1}_n'R^{-1}\mathbf{1}_n} \mathbf{1}_n' + r'\right] R^{-1}Y,
\end{eqnarray}
with mean squared error
\begin{eqnarray}\label{shat}
   s^2(x^*) &=&  \sigma^2_z(1 - 2C'r + C'RC) \\ \nonumber
   &=& \sigma^2_z \left( 1 - r'R^{-1}r +
   \frac{(1-{\bf1_n'}R^{-1}r)^2}{{\bf1_n'}R^{-1}{\bf1_n}}\right),
\end{eqnarray}
where  $r=(r_1(x^*),...,r_n(x^*))'$, $r_i(x^*)=
\text{corr}(z(x^*),z(x_i))$, and $C$ is such that $\hat{y}(x^*) = C'Y$. In practice, the parameters $\mu, \sigma^2_z$ and $\theta$ are replaced with estimates (see Sacks et al. 1989, Santner, Williams and Notz 2003, for details).

\subsection{A computational issue in model fitting}\label{sec:issues}
Fitting a GP model (\ref{GASP})--(\ref{shat}) to a data set with $n$ observations in $d$-dimensional input space requires numerous evaluations of the log-likelihood function for several realizations of the parameter vector $(\theta_1, ..., \theta_d; \mu, \sigma^2_z)$. The closed form estimators of $\mu$ and $\sigma^2_z$, given by
\begin{equation}\label{muhat}
   \hat{\mu}(\theta) = {({\bf1_n}'R^{-1}{\bf1_n})}^{-1}({\bf1_n}'R^{-1}Y)
   \ \text{and} \   \hat{\sigma}^2_z(\theta) =
   \frac{(Y-{\bf1_n}\hat{\mu}(\theta))'R^{-1}(Y-{\bf1_n}\hat{\mu}(\theta))}{n},
\end{equation}
are often used to obtain the profile log-likelihood
\begin{equation}\label{profile-likelihood}
  -2\log L_p \propto \log(|R|) + n \log[(Y-{\bf1_n}\hat{\mu}(\theta))' R^{-1}(Y-{\bf1_n}\hat{\mu}(\theta))],
\end{equation}
for estimating the hyper-parameters $\theta = (\theta_1, ..., \theta_d)$, where $|R|$ denotes the determinant of $R$. Recall from (\ref{corr}), that the correlation matrix $R$ depends on $\theta$ and the design points.

An $n \times n$ matrix $R$ is said to be near-singular (or, ill-conditioned) if its condition number $\kappa(R) = \|R\|\cdot\|R^{-1}\|$ is too large (see Section~4 for details on ``how large is large?"), where $\|\cdot\|$ denotes a matrix norm (we will use the $L_2$--norm). Although these correlation matrices are positive definite by definition, computation of $|R|$ and $R^{-1}$ can sometimes be unstable due to ill-conditioning. This prohibits precise computation of the likelihood and hence the parameter estimates.

Ill-conditioning of $R$ often occurs if any pair of design points are very close in the input space, or $\theta_k$'s are close to zero, i.e., $\sum_{k=1}^d\theta_k|x_{ik}-x_{jk}|^{p_k} \approx 0$. The distances between neighboring points in space-filling designs with large $n$ (sample size) and small $d$ (input dimension) can be very small. Near-singularity is more common in the sequential design setup (e.g., expected improvement based designs, see Jones et al. 1998; Schonlau et al. 1998; Oakley 2004; Huang et al. 2006; Ranjan et al. 2008; Taddy et al. 2009), where the follow-up points tend to ``pile up" near the pre-specified features of interest like the global maximum, contours, quantiles, and so on.

\subsection{The popular approach}\label{sec:popular}
A popular approach to overcome the ill-conditioning of $R$ is to introduce a nugget, $0< \delta < 1$ in the model, and replace the ill-conditioned $R$ with a well-conditioned $R_{\delta} = R + \delta I$ that has a smaller condition number (see Section~4 for details) as compared to that of $R$. Equivalently, one can introduce an independent white noise process in the model
$$ y(x_i) = \mu + z(x_i) + \epsilon_i,\qquad i=1,...,n,$$
where $\epsilon_i$ are i.i.d. $N(0, \sigma^2_{\epsilon})$. That is, $Var(Y) = V(D_0|y(D_0)) = \sigma^2_z R + \sigma^2_{\epsilon} I = \sigma^2_z(R + \delta I)$ for $\delta = \sigma^2_{\epsilon}/\sigma^2_z$. The value of the nugget is bounded above, $\delta < 1$, to ensure that the numerical uncertainty is smaller than the process uncertainty. The resulting BLUP is given by
\begin{equation}\label{eq:yhat_delta}
\hat{y}_{\delta}(x) = \left[ \frac{(1-r'(R+\delta I)^{-1}\mathbf{1}_n)}{\mathbf{1}_n'(R+\delta I)^{-1}\mathbf{1}_n} \mathbf{1}_n' + r'\right] (R+ \delta I)^{-1}Y,
\end{equation}
and the associated mean squared error $s^2_{\delta}(x)$ is
\begin{equation}\label{eq:shat_delta}
 s^2_{\delta}(x) =  \sigma^2_z \left( 1 - 2C_{\delta}'r + C_{\delta}'RC_{\delta} \right),
\end{equation}
where $C_{\delta}$ is such that $\hat{y}_{\delta}(x) = C_{\delta}'Y$.

Theoretically, it is straightforward to see that the use of a positive nugget in the GP model produces a non-interpolator. Jones et al.\ (1998) show that the GP fit given by (\ref{yhat}) and (\ref{shat}) is an interpolator because for $1 \le j \le n$, $r'R^{-1} = e_j'$, where $e_j$ is the $j$-th unit vector, $r = (r_1(x_j),...,r_n(x_j))'$ and $r_i(x_j) = \text{corr}(z(x_i), z(x_j))$. If we use a $\delta\, (>0)$ in the model (i.e., replace $R$ with $R_{\delta}$), then $r'R^{-1}_{\delta} \ne e_j'$ and thus $\hat{y}(x_j) \ne y_j$ and $\hat{s}^2(x_j) \ne 0$. From a practitioner's viewpoint, one could sacrifice exact interpolation if the interpolation accuracy of the fit is within the desired tolerance, but it is not always achievable (see Section~\ref{sec:examples} for illustrations).

The nugget parameter $\delta$ is often estimated along with the other model parameters. However, one of the major concerns in the optimization is that the likelihood (modified by replacing $R$ with $R_{\delta}$) computation fails if the candidate nugget $\delta \in (0, 1)$ is not large enough to overcome ill-conditioning of $R_{\delta}$. To avoid this problem in the optimization, it is common to fix an \emph{ad-hoc} boundary value on the nugget parameter. The resulting maximum likelihood estimate is often close to this boundary value and the fit is not an interpolator of the observed data (i.e., the interpolation error is more than the desired tolerance). Even if the estimated nugget is not near the boundary, the use of a nugget in the model in this manner may introduce unnecessary over-smoothing from a practical standpoint (Section~\ref{sec:examples} presents several illustrations). In the next section, we propose a lower bound on the nugget that minimizes the unnecessary over-smoothing.

\section{Choosing the Nugget}\label{sec:nugget}
Recall from Section~3.2 that an $n \times n$ matrix $R$ is said to be ill-conditioned or
near-singular if its condition number $\kappa(R)$ is too large. Thus, we intend to find $\delta$ such that $\kappa(R_{\delta})$ is smaller than a certain threshold. Our main objectives here are to compute the condition number of $R_{\delta}$ and the threshold that classifies $R_{\delta}$ as well-behaved.

Let $\lambda_1 \le \lambda_2 \le \cdots \le \lambda_n$ be the eigenvalues of $R$. Then, in the $L_2$--norm, $\kappa(R) = \lambda_n/ \lambda_1$  (Golub and Van Loan 1996).
The addition of $\delta$ along the main diagonal of $R$ shifts all of the eigenvalues of $R$ by $\delta$. That is, the eigenvalues of $R_{\delta} = R + \delta I$ are $\lambda_i + \delta$, $i=1,...,n$, where $\lambda_i$ is the $i$-th smallest eigenvalue of $R$. Thus, $R_{\delta}$ is well-conditioned if
 \begin{eqnarray*}
  \log(\kappa(R_{\delta})) & \lessapprox & a\\
  \frac{\lambda_n + \delta}{\lambda_1 + \delta} & \lessapprox & e^{a}\\
  \delta & \gtrapprox & \frac{\lambda_n(\kappa(R)-e^{a})}{\kappa(R)(e^{a}-1)} = \delta_{lb},
 \end{eqnarray*}
where $\kappa(R) = \lambda_n/\lambda_1$ and $e^a$ is the desired threshold for $\kappa(R_{\delta})$. Note that $\delta_{lb}$ is a function of the design points and the hyper-parameter $\theta$.

The closed form expressions for the eigenvalues and hence the condition number of a Gaussian correlation matrix $R$, in (\ref{corr}), for arbitrary $\theta$ and design $\{x_1,...,x_n\}$ is, to our knowledge, yet unknown. If $x \in (-\infty, \infty)^d$ and $x_k \sim N(0, \sigma^2_x)$, closed form expressions of the expected eigenvalues of $R$ are known (see Section 4.3 in Rasmussen and Williams 2006). In our case, $x \in [0,1]^d$, and the design points are often chosen using a space-filling criterion (e.g., Latin hypercube with properties like maximin distance, minimum correlation, OA; uniform designs, and so on). In such cases, one may assume, at most,  $x_k \sim U(0,1)$ for $k=1,...,d$. In fact, the objectives of building efficient emulators for computer simulators often include estimating pre-specified process features of interest, and sequential designs (e.g., expected improvement based designs) are preferred to achieve such goals. In such designs, the follow-up points tend to ``pile up" near the feature of interest. The distributions of such design points are not uniform and can be non-trivial to represent in analytical expressions.
Hence, it is almost surely infeasible to obtain closed form expressions for the eigenvalues of such $R$ in general. Of course, one can compute these quantities numerically. We use Matlab's built-in function \emph{eig} to compute the maximum eigenvalue of $R$ and \emph{cond} to calculate the condition number $\kappa(R)=\lambda_n/\lambda_1$ in the expression of $\delta_{lb}$.

Another important component of the proposed lower bound is the threshold for getting well-behaved non-singular correlation matrices. As one would suspect, the near-singularity of such a correlation matrix depends on $n$, $d$, the distribution of $\{x_1,...,x_n\} \in [0, 1]^d$ and $\theta \in (0, \infty)^d$. We now present the key steps of the simulation algorithm used for estimating the threshold under a specific design framework. The results are averaged over the distribution of $\{x_1,...,x_n\}$ and $\theta$, and thus it is sufficient to find the threshold of $\kappa(R)$.

\begin{figure}[h!]\centering
\includegraphics[height=3.0in,width=3.20in]{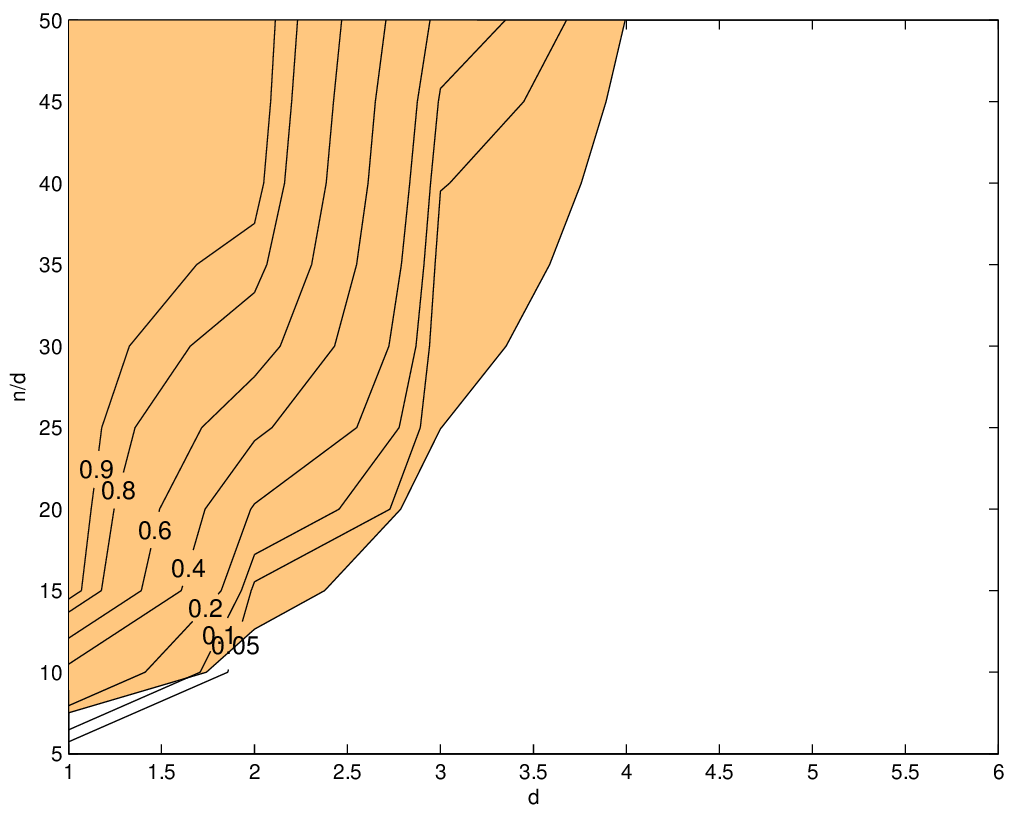}
\includegraphics[height=3.0in,width=3.20in]{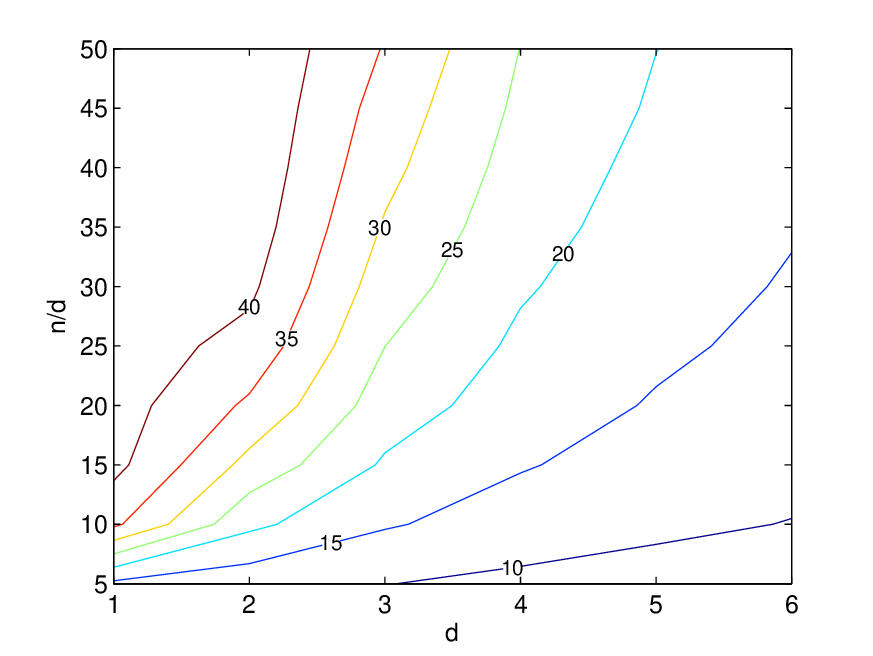}
\vspace{-0.65cm}
\caption{The contours in the left panel show the proportion of correlation matrices flagged as near-singular. The contours in the right panel display average $\log(\kappa(R))$ values. The shaded region in the left panel corresponds to $\log(\kappa(R)) > 25$. }\label{fig:nearsingular_lhs}
\end{figure}

For several combinations of $n$ and $d$, we generate $5000$ correlation matrices where the design points $\{x_1,...,x_n\}$ follow the maximin Latin hypercube sampling scheme (Stein 1987) and $\theta_k$'s are chosen from an exponential distribution with mean $1$. Recall from Section~3.2 that a near-singular (or ill-conditioned) correlation matrix has a large condition number, and $\kappa(R)$ is inversely proportional to $\theta$. Consequently, we focussed on small values of $\theta$ in simulating $R$. These correlation matrices are used to compute the proportion of matrices that are near-singular (see the contours in the left panel of Figure~\ref{fig:nearsingular_lhs}). We used Matlab's built-in function \emph{lhsdesign} to generate the design points and \emph{chol} (which computes the Cholesky factorization) to check whether or not a matrix $R$ was near-singular under the working precision.

\begin{center}
\mbox{\parbox{14cm}{\emph{For a positive definite well-behaved matrix $R$, ``$[U,p] = chol(R)$" produces an upper triangular matrix $U$ satisfying $U'U=R$ and $p$ is zero. If $R$ is not positive definite, then $p$ is a positive integer.}}}\\
\end{center}
We also computed the condition numbers of these $5000$ correlation matrices (using Matlab's built-in function \emph{cond}). The right panel of Figure~\ref{fig:nearsingular_lhs} presents the contours of the average of $\log(\kappa(R))$ for different combinations of $n$ and $d$.

From Figure~\ref{fig:nearsingular_lhs}, it is clear that $a\approx25$ can be used as the threshold for $\log(\kappa(R_{\delta}))$ of a well-behaved correlation matrix $R_{\delta}$. Also note that the proportion of near-singular cases, denoted by the contours in the left panel of Figure~\ref{fig:nearsingular_lhs}, decreases rapidly with the increment in the input dimension. This is somewhat intuitive because the volume of the void (or unexplored region) increases exponentially with the dimension, and a really large space-filling design is needed to jeopardize the conditioning of the correlation matrices in high dimensional input space. For other design schemes (e.g., sequential designs), one can follow these steps to estimate the threshold for the condition number of well-behaved correlation matrices.

The lower bound on the nugget is only a sufficient condition and not a necessary one for $R_{\delta}$ to be well-conditioned. For instance, a correlation matrix with 100 design points in $(0,1)^2$ chosen using a space-filling criterion may lead to a well-behaved $R$ if $\theta$ is very large. If the correlation matrix is well-conditioned, $R$ should be used instead of $R_{\delta}$, i.e.,
 \begin{equation}
 \delta_{lb} = \max\left\{ \frac{\lambda_n(\kappa(R)-e^{a})}{\kappa(R)(e^{a}-1)}, 0 \right\}.
 \end{equation}\label{delta_lb}
That is, when $R$ is well-behaved our approach allows $\delta_{lb}$ to be zero and hence a more accurate surrogate can be obtained as compared to the popular approach (Section~3.3) where a non-zero nugget is forced in the model which may lead to undesirable over-smoothing. This could be of concern in high dimensional input space, because the proportion of near-singular cases decreases with the increment in the input dimension. Example~3 demonstrates the performance of the proposed methodology over the popular approach for an eight-dimensional simulator.

Although the use of $\delta_{lb}$ in the GP model minimizes the over-smoothing, $\delta_{lb}$ may not be small enough to achieve the desired interpolation accuracy (see Examples~1 and 2 for illustrations), and choosing $\delta <\delta_{lb}$ may lead to ill-conditioned $R$. This may not be a big issue if one believes that the simulator is somewhat noisy and/or the statistical emulator is biased due to mis-specification in the correlation structure or model assumptions. In such cases, a little smoothing might be a good idea. However, controlling the amount of smoothing is a non-trivial task and requires more attention. On the other hand, over-smoothing is undesirable if the experimenter believes that the computer simulator is deterministic and the statistician is confident about the choice of the emulator (we consider the GP model with Gaussian correlation structure). Under these assumptions, we now propose a new predictor that can achieve the desired level of interpolation accuracy.

\section{New Iterative Approach}\label{sec:iterative}

In this section, we propose a predictor that is based on the iterative use of a nugget $\delta \in (0, 1)$. This approach does not depend severely on the magnitude of the nugget, and the results developed here are based on an arbitrary $0<\delta<1$, large enough to ensure $R_{\delta}$ well-behaved. However, choosing $\delta > \delta_{lb}$ may require more iterations to attain the desired interpolation accuracy, and we recommend using $\delta_{lb}$. We also show that the proposed predictor converges to the interpolator (\ref{yhat}) and (\ref{shat}).

Recall that the key problem here is the inaccurate computation of $|R|$ and $R^{-1}$ due to ill-conditioning of $R$. The main idea of the new approach is to rewrite the profile log-likelihood as
\begin{equation}\label{new-profile-likelihood}
  -2\log L_p \propto -\log(|R^{-1}|) + n \log[(Y-{\bf1_n}\hat{\mu}(\theta))' R^{-1}(Y-{\bf1_n}\hat{\mu}(\theta))],
\end{equation}
and replace the ill-conditioned $R^{-1}$ with a well-behaved quantity. This modified profile log-likelihood can then be optimized to get the parameter estimates. Next, we describe how to find the appropriate well-behaved substitute for $R^{-1}$.

In the same spirit as the popular approach, we attempt to evaluate
$R^{-1}w$ by solving $Rt = w$, under the assumption that $R$ cannot be inverted accurately (i.e., $R$ is near-singular) and there exists a $\delta \in (0,1)$ such that $R_{\delta} = (R + \delta I)$ is well-conditioned. In an attempt to find an interpolator of the simulator (up to certain accuracy), our objective is to find $t^* = f(R_{\delta}, w)$ that is a better approximation of $t=R^{-1}w$ as compared to $\tilde{t} = R_{\delta}^{-1}w$, suggested by the popular approach. To achieve this goal, we propose to use \emph{iterative regularization} (e.g., Tikhonov 1963, Neumaier 1998), a technique for solving ill-conditioned systems of equations.

Let $s_0 = w$ and $s_i, i=1,...,M$, be a sequence of vectors obtained by recursively solving the system of equations given by
\begin{equation}\label{eq:iter_s}
(R+\delta I) s_i = \delta s_{i-1}.
\end{equation}
Then, the estimate of $t = R^{-1}w$ after the $i$-th iteration ($1 \le i \le M$) of regularization is given by
\begin{equation}\label{eq:iter_t}
t_i = t_{i-1} + \frac{s_i}{\delta},
\end{equation}
where $t_0$ is a vector of zeros. The final solution with $M$ iterations of regularization,
$$ t_M = \sum_{k=1}^M \delta^{k-1}(R+\delta I)^{-k} w, $$
requires only one direct inversion (or one Cholesky decomposition) of $R_{\delta} = R + \delta I$, followed by $M$ forward and backward substitutions. The proposed approximation of $t = R^{-1}w$ is $t_M$,  with $M \ge 1$ chosen to satisfy the interpolation accuracy requirement. Lemma~\ref{lemma:taylor} shows that the iterative regularization approach in (\ref{eq:iter_s}) and (\ref{eq:iter_t}) leads to a solution that is a generalization of the popular approach outlined in Section~3.3.

\begin{lemma}\label{lemma:taylor}
Let $R$ be a $n \times n$ positive definite correlation matrix, $I$ be the $n \times n$ identity matrix, and $0< \delta <1$ be a constant, then
$$ R^{-1} = \sum_{k=1}^{\infty} \delta^{k-1}(R+\delta I)^{-k}.$$
\end{lemma}

The convergence of this infinite series follows from the von Neumann series (the matrix version of the Taylor series, Lebedev 1997) expansion of $g(u)= (R + u I)^{-1}$ around $u=\delta$:
\begin{eqnarray*}
g(u) &=& g(\delta) + (u-\delta)g'(\delta) + \frac{(u-\delta)^2}{2!}g''(\delta) + \frac{(u-\delta)^3}{3!}g'''(\delta) + \cdots,\\
\textrm{i.e.,} \quad (R+uI)^{-1} &=& (R+\delta I)^{-1} + (u-\delta)(-1)(R+\delta I)^{-2} \\
 && + \ \frac{(u-\delta)^2}{2!}(-1)(-2)(R+\delta I)^{-3} + \cdots \\
 &=& (R+\delta I)^{-1} - (u-\delta)(R+\delta I)^{-2} +
  (u-\delta)^2(R+\delta I)^{-3} - \cdots.
\end{eqnarray*}
Setting $u=0$, we get $R^{-1} = \sum_{k=1}^{\infty} \delta^{k-1}(R+\delta I)^{-k}$ and thus the proposed solution obtained using the iterative regularization is the $M$-th order von Neumann approximation of $R^{-1}$. The predictor $\hat{y}_{\delta}$ in the popular approach (\ref{eq:yhat_delta}) uses $t_1$, the first order von Neumann approximation, and hence our proposed approach is a generalization of the popular approach.

The proposed regularization is implemented by optimizing the modified profile log-likelihood
\begin{equation}\label{final-profile-likelihood}
  -2\log L_p \propto -\log(|R^{-1}_{\delta, M}|) + n \log[(Y-{\bf1_n}\hat{\mu}(\theta))' R^{-1}_{\delta, M}(Y-{\bf1_n}\hat{\mu}(\theta))],
\end{equation}
where $R^{-1}_{\delta, M} = \sum_{k=1}^{M} \delta^{k-1}(R+\delta I)^{-k}$. Closed form expressions for $\hat{\mu}(\theta)$ and $\hat{\sigma}^2_z(\theta)$ are the same as in (\ref{muhat}) subject to $R^{-1}$ replaced by $R^{-1}_{\delta, M}$. The new regularized predictor $\hat{y}_{\delta, M}(x)$ at $x \in \chi$ is

\begin{equation}\label{eq:yhat_new}
   \hat{y}_{\delta, M}(x)
   = \left[\frac{\left(1-r'R^{-1}_{\delta, M}\mathbf{1}_n\right)}{\left(\mathbf{1}_n'R^{-1}_{\delta, M}\mathbf{1}_n\right)}  \mathbf{1}_n' + r' \right] R^{-1}_{\delta, M}Y,
\end{equation}
and the corresponding MSE $s^2_{\delta, M}(x)$ is given by
\begin{equation}\label{eq:shat_new}
   s_{\delta, M}^2(x)
   = \sigma^2_z(1  -2C_{\delta,M}'r + C_{\delta,M}'RC_{\delta,M}),
\end{equation}
where $C_{\delta,M}$ is such that $\hat{y}_{\delta, M}(x) = C_{\delta,M}'Y$. Lemmas~\ref{lemma:1} and \ref{lemma:2} establish the convergence results for an arbitrary $0< \delta < 1$.

\begin{lemma}\label{lemma:1}
Let $R$ be a near-singular correlation matrix as defined in (\ref{corr}), and $0< \delta <1$ be a nugget such that $R + \delta I$ is well-behaved. Then, for every $x^* \in \chi = [0, 1]^d$,
$$ \lim_{M \rightarrow \infty} \hat{y}_{\delta, M}(x^*) = \hat{y}(x^*),$$
where $\hat{y}(x^*)$ and $\hat{y}_{\delta, M}(x^*)$ are defined in (\ref{yhat}) and (\ref{eq:yhat_new}) respectively.
\end{lemma}

The proof follows from Lemma~\ref{lemma:taylor} and using $\underset{M \rightarrow \infty}{\lim} R^{-1}_{\delta, M} = R^{-1}$ in (\ref{eq:yhat_new}). It is straightforward to show that $C_{\delta, M}$ in (\ref{eq:shat_new}) converges to $C$ in (\ref{shat}) as $M \rightarrow \infty$. This also proves the next result on the convergence of the mean squared error for the proposed predictor.

\begin{lemma}\label{lemma:2}
Let $R$ be a near-singular correlation matrix as defined in (\ref{corr}), and $0< \delta <1$ be a nugget such that $R + \delta I$ is well-behaved. Then, for every $x^* \in \chi=[0, 1]^d$,
$$ \lim_{M \rightarrow \infty} s^2_{\delta, M}(x^*) = s^2(x^*),$$
where $s(x^*)$ and $s_{\delta, M}(x^*)$ are defined in (\ref{shat}) and (\ref{eq:shat_new}) respectively.
\end{lemma}

Lemmas~\ref{lemma:1} and \ref{lemma:2} prove that even if a few pairs of points are too close
together in the input space, or $\theta_k$'s are close to zero to cause near-singularity of $R$, the proposed iterative predictor converges to an interpolator as $M$ increases (i.e., for $1 \le i \le n$, $\hat{y}_{\delta, M}(x_i) \rightarrow y_i$ and $s^2_{\delta, M}(x_i) \rightarrow 0$ as $M \rightarrow \infty$).\\[-0.2cm]

\noindent \textbf{Remark:}  In practice, when a pre-specified interpolation accuracy is desired, the proposed iterative approach suggests refitting the GP model (i.e., optimization of (\ref{final-profile-likelihood})) for different choices of $M\ge 1$. Note that the parameter estimates change with $M$ which allows for the extra flexibility in the model that adjusts the over-smoothed portion of the surrogate. First of all, the computational cost of fitting this model increases with $M$. Secondly, the combined cost of refitting the model for different values of $M$ can be quite large. Although the numerical stability in computing $R^{-1}_{\delta, M}$ does not change with $M$, computation of $|R^{-1}_{\delta, M}|$ can become less numerically stable with increasing $M$. This is because $R^{-1}_{\delta, M} \rightarrow R^{-1}$ as $M \rightarrow \infty$ and the computation of $|R^{-1}|$ is assumed to be unstable. Considering these issues, we recommend optimizing the profile log-likelihood (\ref{final-profile-likelihood}) with $M=1$ to obtain $\hat{\theta}_{mle}$ and $\delta_{lb}(\hat{\theta}_{mle})$, and then use it to compute $\hat{y}_{\delta, M}(x)$ and $\hat{s}_{\delta, M}(x)$ for any $M \ge 1$ by following the iterative regularization steps outlined above. \\

The convergence results in Lemmas~\ref{lemma:taylor}, \ref{lemma:1} and \ref{lemma:2} do not depend on the choice of $\theta$ and $\delta$ in $R_{\delta} = R + \delta I$, and so the predictor obtained is still an interpolator. The key steps required for the implementation of the proposed approach are as follows:
\begin{enumerate}
 \item Computation of the profile log-likelihood (\ref{new-profile-likelihood}) for the estimation of $\theta$.
 \begin{enumerate}
   \item Choose a candidate $\theta$ in $\Theta^d$ and compute $R$.
   \item Compute the lower bound of nugget $\delta_{lb}$ in (9). Note that $\delta_{lb}$ is a function of the hyper-parameters $\theta$, the design matrix and the threshold.
   \item Replace $R^{-1}$ with $R_{\delta_{lb},1}^{-1}$ in the likelihood (\ref{new-profile-likelihood}).
 \end{enumerate}
 \item Obtain the parameter estimates $\hat{\theta}$ and $\delta_{lb}(\hat{\theta})$ by optimizing the profile log-likelihood. Then compute $\hat{\mu}(\hat{\theta})$ and $\hat{\sigma}^2_z(\hat{\theta})$.

 \item Use the parameter estimates $\hat{\theta}$, $\delta_{lb}(\hat{\theta})$, $\hat{\mu}(\hat{\theta})$ and $\hat{\sigma}^2_z(\hat{\theta})$ to compute the regularized emulator given by $\hat{y}_{\delta_{lb},M}(x)$ and $\hat{s}^2_{\delta_{lb},M}(x)$ in (\ref{eq:yhat_new}) and (\ref{eq:shat_new}) respectively.
\end{enumerate}
The number of iterations $(M)$ in $\hat{y}_{\delta_{lb},M}(x)$ and $\hat{s}^2_{\delta_{lb},M}(x)$ depends on the desired interpolation accuracy, and one can build stopping rules for attaining the pre-specified accuracy in (\ref{eq:yhat_new}). We use Mahalanobis distance (Bastos and O'Hagan 2009) to compute the accuracy of the predictor. The interpolation accuracy is measured by
$$ \xi_{I,k}^0 = \log_{10} \left[(y(D_0) - \hat{y}_{\delta_{lb},k}(D_0))'\{V(D_0|y(D_0))\}^{-1}(y(D_0) - \hat{y}_{\delta_{lb},k}(D_0))\right], $$
where $\hat{y}_{\delta_{lb},k}(D_0) = (\hat{y}_{\delta_{lb},k}(x_1),...,\hat{y}_{\delta_{lb},k}(x_n))'$, and $V(D_0|y(D_0)) = \sigma^2_z(R+\delta_{lb}I)$. Similarly,
$$ \xi_{I,k} = \log_{10} \left[(\hat{y}_{\delta_{lb},k}(D_0) - \hat{y}_{\delta_{lb},k-1}(D_0))' \{V(D_0|y(D_0))\}^{-1}(\hat{y}_{\delta_{lb},k-1}(D_0) - \hat{y}_{\delta_{lb},k}(D_0))\right], $$
measures the improvement of the predictor $\hat{y}_{\delta, k}$ in interpolating the data by increasing the number of terms in the von Neumann approximation.
Lemmas~\ref{lemma:1} and \ref{lemma:2} show that both $\xi_{I,k}$ and $\xi_{I,k}^0$ tend to $-\infty$ as $k$ increases. As $\xi_{I,k}$ tends to $-\infty,$ the predictor $\hat{y}_{\delta,k}$ is stabilizing. While as $\xi_{I,k}^0$  tends to $-\infty,$  the predictor in (\ref{eq:yhat_new}) is converging to the BLUP in (\ref{yhat}). As we will see in Example~1, the rates of convergence of $\xi_{I,k}$ and $\xi_{I,k}^0$ may differ. That is, both of these measures ($\xi_{I,k}^0$ and $\xi_{I,k}$) can be used in practice to choose appropriate $M$ for achieving the desired interpolation accuracy. For measuring the prediction accuracy (at out-of-sample points), we define an analogous quantity
$$ \xi_{P,k}^0 = \log_{10} \left[(y(D_{new}) - \hat{y}_{\delta_{lb},k}(D_{new}))'\{V(D_{new}|y(D_0))\}^{-1}(y(D_{new}) - \hat{y}_{\delta_{lb},k}(D_{new}))\right], $$
where $D_{new}$ is a set of $n_{new}$ unsampled points in the input space. The parameters $\sigma_z^2$ and $\theta$ in the covariance matrix $V(D_{new}|y(D_0)) = \sigma^2_z(R+\delta_{lb}I)$ are estimated from the original data $D_0$, but $\delta_{lb}$ was recomputed for $D_{new}$.  For the simulated examples considered in this paper, we used maximin Latin hypercube designs of size $n_{new} = 1000\cdot d$ as $D_{new}$, whereas the tidal power application used a holdout set for $D_{new}$. The next section illustrates that even the best choice of $\delta$ can lead to over-smooth emulators, and the iterative approach is advantageous.

\section{Examples}\label{sec:examples}
To illustrate the proposed approach we first present a few simulated examples. The performance of the new iterative predictor is also compared with the popular approach. Then, we revisit the tidal power modeling example.

\begin{figure}[h!]\centering
\subfigure[Popular fit with MLE]{\label{fig:eg2a} \includegraphics[height=2.0in,width=3.10in]{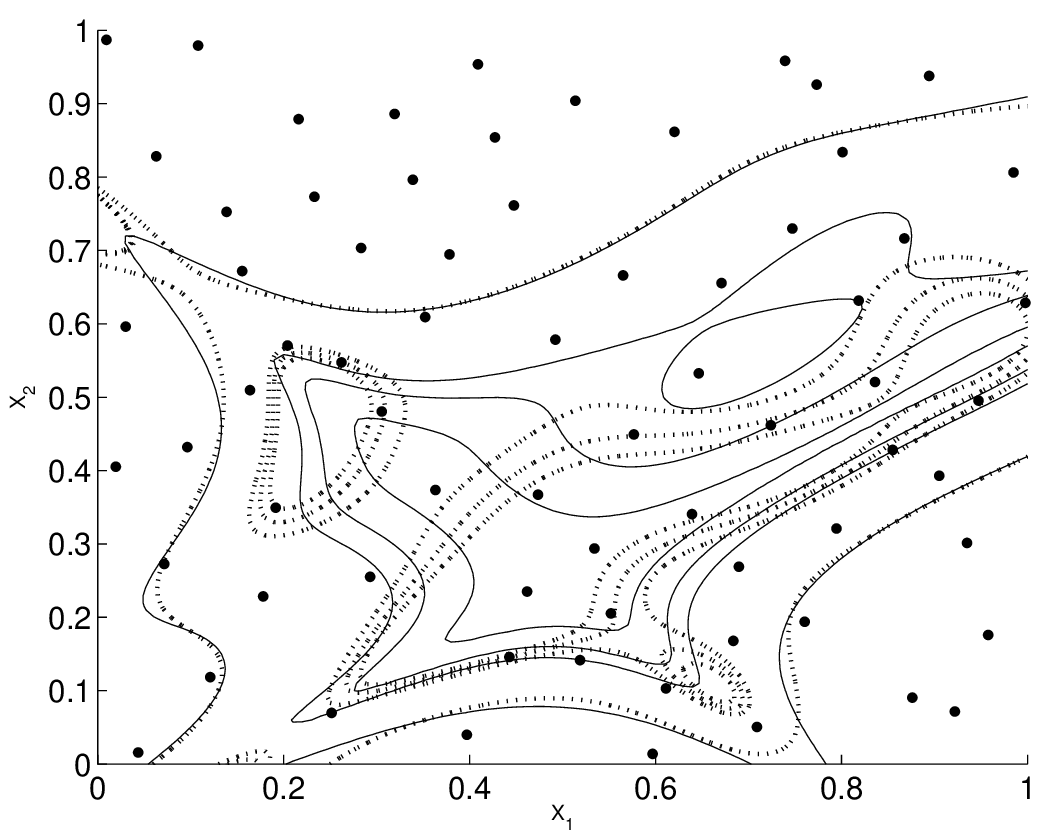}}
\subfigure[$M=1$ (with lower bound)]{\label{fig:eg2b} \includegraphics[height=2.0in,width=3.10in]{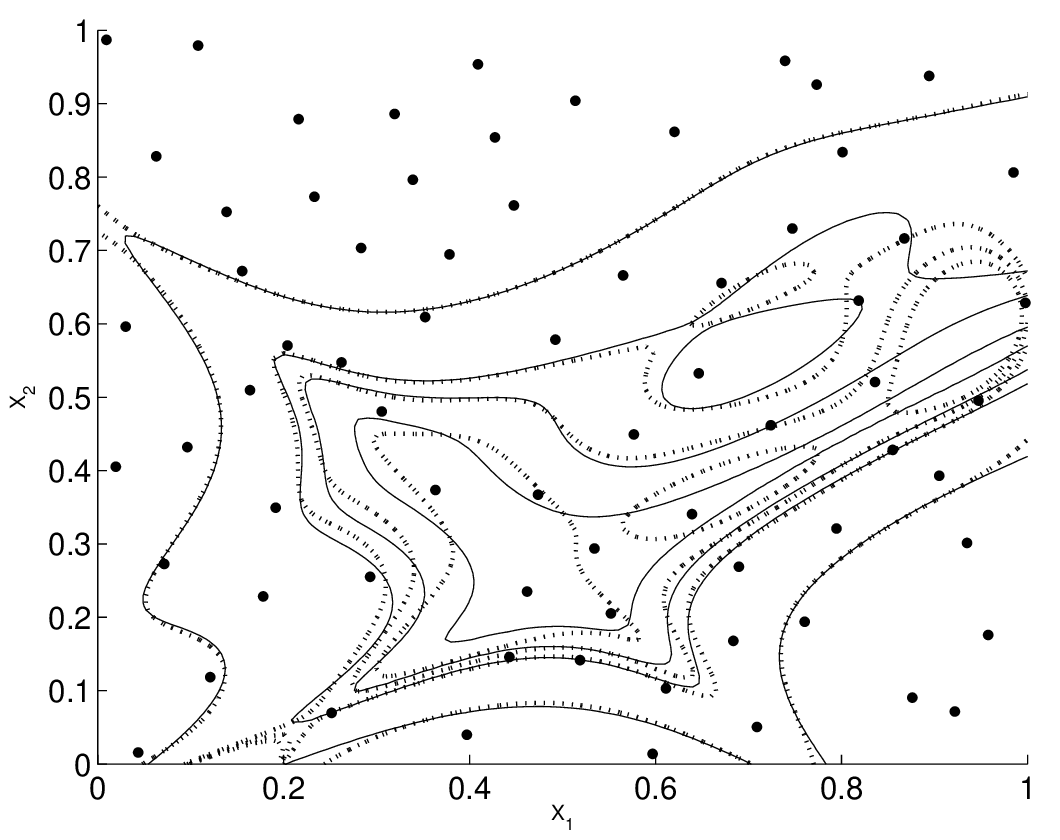}}
\subfigure[$M=5$ (with lower bound)]{\label{fig:eg2c} \includegraphics[height=2.0in,width=3.10in]{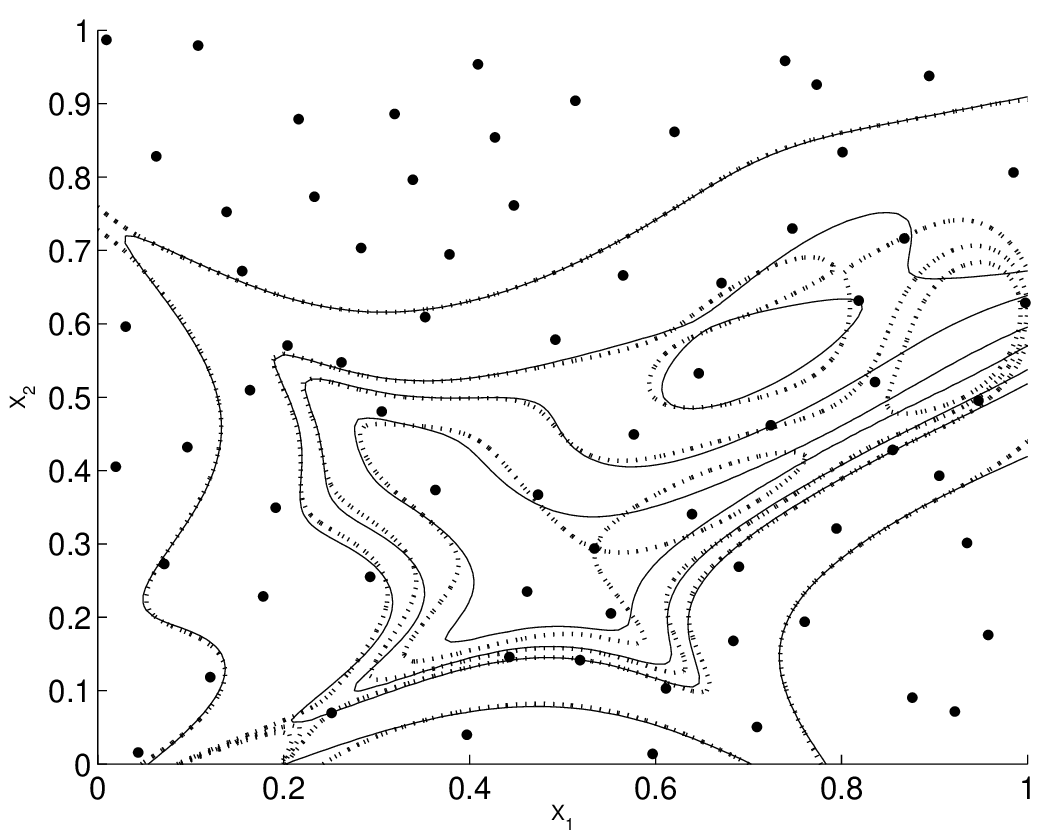}}
\subfigure[$M=20$ (with lower bound)]{\label{fig:eg2d} \includegraphics[height=2.0in,width=3.10in]{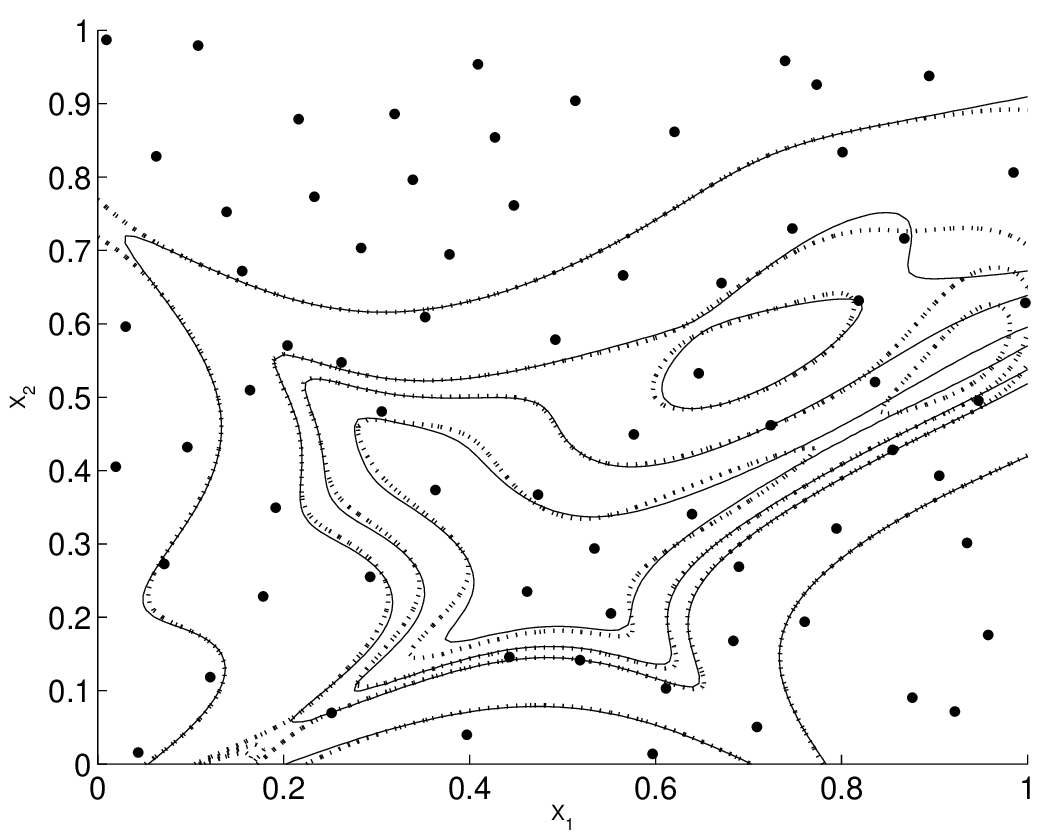}}
\caption{The dots denote the design points, the solid curves denote the contours of the true GoldPrice function, and the dashed curves represent the contours of the predicted surfaces.}\label{fig:eg2_iter}
\end{figure}

\begin{figure}[h!]\centering
  \includegraphics[height=2.25in,width=4.5in]{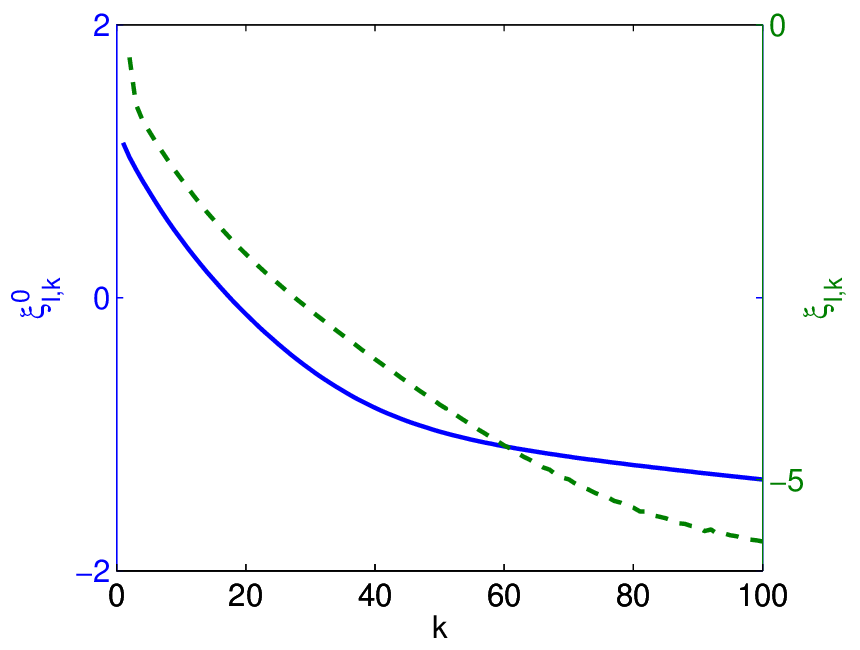} \\[-0.75cm]
\caption{Convergence of $\xi_{I,k}$ (dashed curve - right axis) and $\xi_{I,k}^0$ (solid curve - left axis).}\label{fig:eg2_conv}
\end{figure}

\begin{example}{\rm
Let $x_1, x_2 \in [0, 1]$, and the underlying deterministic simulator output be generated using the GoldPrice function (Andre, Siarry and Dognon 2000),

{\footnotesize
\begin{equation*}
  f(x_1, x_2) =
  \left[1 + \left(\frac{x_1}{4}+2+\frac{x_2}{4}\right)^2\left\{5
 - \frac{7x_1}{2} + 3\left(\frac{x_1}{4} +\frac{1}{2}\right)^2 -\frac{7x_2}{2} +
  \left(\frac{3x_1}{2} +3 \right) \left(\frac{x_2}{4} + \frac{1}{2}\right) + 3
  \left(\frac{x_2}{4} +\frac{1}{2}\right)^2\right\}\right]*
\end{equation*}
\begin{equation*}
    \left[30+\left(\frac{x_1}{2}-\frac{1}{2}-\frac{3x_2}{4}\right)^2
    \left\{26-8x_1+12\left(\frac{x_1}{4}+\frac{1}{2}\right)^2+12x_2-(9x_1+18)
    \left(\frac{x_2}{4}+\frac{1}{2}\right)+27\left(\frac{x_2}{4}+
    \frac{1}{2}\right)^2\right\}\right].
\end{equation*}  }

For illustration purposes, we intentionally select a maximin Latin hypercube design (Stein 1987) with $n=70$ points that leads to an ill-conditioned correlation matrix for small $\theta \in (0,\infty)^2$. It turns out that for this particular design (see Figure~3), the correlation matrix $R$ is ill-conditioned if $\theta_1\cdot\theta_2 \lessapprox 3$. Figure~\ref{fig:eg2_iter}(a) presents the contours (at heights $y = 120, 500, 1000$ and $10000$) of the true simulator (solid curve) and the GP surrogate fit (obtained using the methodology outlined in Sections~3.1 and 3.3). For successful implementation of the popular approach, we optimized the likelihood in the parameter space $\delta \in (10^{-5}, 1)$ and $\theta \in (0, \infty)^2$. The parameter estimates for the GP fit are $\hat{\delta}_{mle} = 1.06\cdot 10^{-5}$ and $\hat{\theta}_{mle} = (5.01, 7.33)$. Note that $\hat{\delta}_{mle}$ is close to the boundary and the fitted surrogate is significantly different than reality in the central part of the input space.

The parameter estimates for the GP model fit obtained from the proposed method are $\hat{\theta}_{mle} = (2.26, 2.75)$ and $\delta_{lb}(\hat{\theta}_{mle}) = 5.26\cdot 10^{-10}$. The GP surrogate for $M=1$, in Figure~\ref{fig:eg2_iter}(b), shows a much better fit, which is further improved by the iterative approach (see Figure~\ref{fig:eg2_iter}(c) and \ref{fig:eg2_iter}(d)). Figure~\ref{fig:eg2_conv} shows that $\xi_{I,k}$ goes to $-\infty$ at a faster rate as compared to $\xi_{I,k}^0$.
}\label{exam:exam2}\end{example}


Table~\ref{tab:goldprice} summarizes the results from a detailed simulation study based on several combinations of the design sizes and the boundary values of $\delta$ in the likelihood optimization. For fair comparison between the two methodologies, first we use the proposed model with only one term in the von Neumann approximation (i.e., $M=1$) and the popular method with $\hat{\delta}_{mle}$. We then increase the number of iterations to measure the improvement in the interpolation accuracy. The results in Table~\ref{tab:goldprice} are summarized over model fits with $50$ random maximin Latin hypercube designs. These designs are generated using Matlab's built-in function \emph{lhsdesign} which takes random starting points and hence the output designs are random. The table entries are $P_{50}\ (P_5, P_{95})$, where $P_r$ denotes the $r$-th percentile of $\xi_{I,M}^0$ values obtained from the model fits.
\begin{table}[!h]\centering \caption{Median and $(P_5, P_{95})$ of $\xi_{I,M}^0$ values for the proposed approach (denoted by ``$lb$'') and the popular approach (denoted by ``$mle$'') applied to the GoldPrice function.}
{\footnotesize
\begin{tabular}{|c|c|c|c|c|}
\hline
                & $n=25$ & $n=50$ & $n=75$ & $n=100$\\
\hline
$\hat{\delta}_{mle} \ge 10^{-5}$  &  0.71 (-3.60, 3.02)& 1.36 (0.61, 1.74)& 2.12 (1.91, 2.49)& 2.43 (2.26, 2.60)\\
$\hat{\delta}_{mle} \ge 10^{-10}$ & -0.89 (-7.08, 2.14)& 1.07 (-3.28, 2.03)& 1.48 (0.46, 2.10)& 1.21 (0.41, 1.70)\\
\hline
$\delta_{lb} (M=1)$          & -25.71 (-28.13, -20.50)& -16.68 (-20.49, -14.40)& 0.85 (0.63, 1.17)& 1.09 (0.90, 1.26)\\
$\delta_{lb} (M=5)$          &  &  & 0.19 (-0.29, 0.70)& 0.43 (0.27, 0.72)\\
$\delta_{lb} (M=20)$         &  &  & -0.48 (-0.91, -0.09)& -0.07 (-0.35, 0.14)\\
\hline
\end{tabular}\label{tab:goldprice}
}
\end{table}

From Table~\ref{tab:goldprice}, it is clear that the interpolation error of the surrogates fitted using the popular approach decreases by lowering the boundary value of the nugget parameter (from $10^{-5}$ to $10^{-10}$) in the optimization problem. It turns out that the correlation matrices are well-behaved for small designs (i.e., $n=25$ and $n=50$) and non zero nuggets are not required for a numerically stable model fitting process. This is captured by the proposed approach, as $\delta_{lb}(\hat{\theta}_{mle}) = 0$ and the interpolation error is much smaller than in the popular approach where a non-zero nugget is forced in the model. Consequently, the iterative approach is not used for these cases. For $n=75$ and $n=100$, the correlation matrices turn out to be near-singular for $\theta$ near $\hat{\theta}_{mle}$, and non-zero $\delta$ had to be used for numerically stable computation. It is clear from the last three rows that the proposed iterative approach leads to improvement in the interpolation accuracy.

Table~\ref{tab:goldprice_pred} summarizes the corresponding prediction accuracy values, $\xi_{P,M}^0$,  where the test set of unsampled points, $D_{new}$, is a randomly chosen $2000$-point maximin Latin hypercube design. The simulation results are based on 50 realizations. As before, the maximin Latin hypercube designs were generated using the built-in function \emph{lhsdesign} in Matlab and the output designs are random. The results illustrate that the proposed iterative approach also improves the prediction at out-of-sample points. Note that the improvement in prediction accuracy is not as significant as the improvement in interpolation accuracy. This is expected as the proposed methodology is geared towards improving the approximation of the interpolator.

\begin{table}[!h]\centering \caption{Median and $(P_5, P_{95})$ of $\xi_{P,M}^0$ values for the predictors in the proposed approach (denoted by ``$lb$'') and the popular approach (denoted by ``$mle$'') applied to the GoldPrice function.}
{\footnotesize
\begin{tabular}{|c|c|c|c|c|}
\hline
                & $n=25$ & $n=50$ & $n=75$ & $n=100$\\
\hline
$\hat{\delta}_{mle} \ge 10^{-5}$  &  4.59 (3.71, 6.17)& 3.45 (3.25, 3.75)& 3.34 (3.18, 3.46)& 3.23 (3.12, 3.43)\\
$\hat{\delta}_{mle} \ge 10^{-10}$ &  4.49 (3.68, 6.01)& 3.40 (3.12, 3.85)& 2.76 (2.50, 3.03)& 2.30 (1.85, 2.52)\\
\hline
$\delta_{lb} (M=1)$          & 4.28 (3.59, 5.72)& 3.22 (3.02, 3.54)& 2.29 (2.07, 2.52)& 1.94 (1.68, 2.18)\\
$\delta_{lb} (M=5)$          &  & & 2.15 (1.90, 2.42)& 1.77 (1.55, 2.02)\\
$\delta_{lb} (M=20)$         &  & & 2.08 (1.71, 2.50)& 1.69 (1.47, 1.98)\\
\hline
\end{tabular}\label{tab:goldprice_pred}
}
\end{table}

\begin{example}{\rm Suppose the deterministic simulator outputs are generated using the three dimensional Perm function (Yang 2010) given by
$$ f(x) = \sum_{k=1}^3 \left[\sum_{i=1}^3 (i^k+\beta)\left((x_i/i)^k - 1\right)\right]^2, $$
where $x = (x_1, x_2, x_3)$ and the $i$-th input variable $x_i \in[-3, 3]$ for $i=1,...,3$. For convenience, we re-scale the input variables in $[0, 1]$. As in the previous example, we fit the GP model using both the popular method (Section~3.3) and the proposed approach (Sections~4 and 5) to the data generated by evaluating the Perm function at $n$ design points. Table~\ref{tab:perm} compares the median and the two tail percentiles $(P_5, P_{95})$ of $\xi_{I,M}^0$ values obtained from fitting GP models to 50 data sets (maximin Latin hypercube designs generated using Matlab's built-in function \emph{lhsdesign}) of different run-sizes. Here also, we pre-specified the boundary values for estimating $\delta$ in the popular approach.

\begin{table}[!h]\centering \caption{Median and $(P_5, P_{95})$ of $\xi_{I,M}^0$ values for the proposed approach (denoted by ``$lb$'') and the popular approach (denoted by ``$mle$'') applied to the perm function.}
{\footnotesize
\begin{tabular}{|c|c|c|c|c|}
\hline
                & $n=25$ & $n=50$ & $n=75$ & $n=100$\\
\hline
$\hat{\delta}_{mle} \ge 10^{-5}$  &  2.42 (-2.78, 4.29)& 1.70 (0.00, 2.17)& 2.79 (1.56, 3.16)& 3.12 (2.22, 3.63)\\
$\hat{\delta}_{mle} \ge 10^{-10}$ &  2.24 (-4.17, 3.85)& 1.69 (-0.36, 2.33)& 2.90 (1.39, 3.28)& 3.07 (1.72, 3.68)\\
\hline
$\delta_{lb} (M=1)$        & -26.46 (-27.53, -24.97)& -21.37 (-22.98, -18.99)& -18.76 (-20.07, -17.35)& 1.76 (-20.35, 1.81)\\
$\delta_{lb} (M=5)$        &  & & & -16.71 (-20.22, 1.73)\\
\hline
\end{tabular}\label{tab:perm}
}
\end{table}

As in Example~1, the interpolation errors of the GP fits obtained through the popular approach are slightly reduced by lowering the boundary value of $\delta$ (from $10^{-5}$ to $10^{-10}$) in the optimization process. The small values of the percentiles of $\xi_{I,M}^0$ in the row labelled ``$\delta_{lb}\, (M=1)$'' of Table~\ref{tab:perm} indicate that the correlation matrices are well-behaved (i.e., $\delta_{lb}(\hat{\theta}_{mle}) = 0$) for most of the designs with runs-sizes $n=25$, $50$ and $75$. It turns out that approximately 46\% of the correlation matrices are well-behaved for designs of size $n=100$. In ``$\delta_{lb}\, (M=5)$" case, the interpolation accuracy has increased and 53\% of the designs of size $n=100$ show $\delta_{lb}(\hat{\theta}_{mle}) =0$. The proposed approach facilitates the inclusion of a non-zero nugget only when required for fixing the ill-conditioning problem. Clearly, the number of realizations that required a non-zero nugget in the GP models here is much smaller than in the GoldPrice example. This is expected because getting near-singular correlation matrices becomes less likely as the dimensionality of the input space increases.

The corresponding prediction accuracy measures for 50 simulations are summarized in Table~\ref{tab:perm_pred}. The test set, $D_{new}$, required for computing $\xi_{P,M}^0$, is a randomly chosen $3000$-point maximin Latin hypercube design. As in Example~1, the prediction accuracy increases with $M$, the number of iterations, and by lowering the boundary value of the nugget in the optimization problem.

\begin{table}[!h]\centering \caption{Median and $(P_5, P_{95})$ of $\xi_{P,M}^0$ values for the predictors in the proposed approach (denoted by ``$lb$'') and the popular approach (denoted by ``$mle$'') applied to the Perm function.}
{\footnotesize
\begin{tabular}{|c|c|c|c|c|}
\hline
                & $n=25$ & $n=50$ & $n=75$ & $n=100$\\
\hline
$\hat{\delta}_{mle} \ge 10^{-5}$  &  6.78 (4.94, 7.43)& 4.24 (4.06, 4.44)& 4.01 (3.91, 4.28)& 4.02 (3.80, 4.16)\\
$\hat{\delta}_{mle} \ge 10^{-10}$ &  6.68 (5.21, 7.22)& 4.23 (4.12, 4.40)& 4.10 (3.90, 4.27)& 3.90 (3.52, 4.24)\\
\hline
$\delta_{lb} (M=1)$        & 5.87 (4.70, 6.88)& 4.08 (3.91, 4.28)& 3.80 (3.60, 4.00)& 2.60 (2.21, 3.89)\\
$\delta_{lb} (M=5)$        & & & & 2.49 (2.17, 3.86)\\
\hline
\end{tabular}\label{tab:perm_pred}
}
\end{table}
}
\end{example}

\begin{example}{\rm The borehole model is a more realistic deterministic simulator, that models the flow rate through a borehole which is drilled from the ground surface through two aquifers, and is commonly used in computer experiments (e.g., Joseph, Hung and Sudjianto 2008) to compare different methods. The flow rate is given by
$$
f(x) = \frac{2\pi T_u (H_u - H_l)}{\log(r/r_w)\left[1+\frac{2LT_u}{\log(r/r_w)r^2_wK_w} + \frac{T_u}{T_l}\right]},
$$
where $x = (r_w, r, T_u, T_l, H_u, H_l, L, K_w)$, and the input $r_w \in [0.05, 0.15]$ is the radius of the borehole, $r \in [100, 50000]$ is the radius of the influence, $T_u \in [63070, 115600]$ is the transmissivity of the upper aquifer, $T_l \in [63.1, 116]$ is the transmissivity of the lower aquifer, $H_u \in [990, 1110]$ is the potentiometric head of the upper aquifer, $H_l \in [700, 820]$ is the potentiometric head of the lower aquifer, $L \in [1120, 1680]$ is the length of the borehole and $K_w \in [9855, 12045]$ is the hydraulic conductivity of the borehole. For convenience, we re-scale the input variables to $[0, 1]$.

Table~\ref{tab:borehole} compares the median and two tail percentiles $(P_5, P_{95})$ of $\xi_{I,M}^0$ values obtained from the GP model surrogates fitted to 50 random maximin Latin hypercube designs via the two methods. Since the simulator is $8$-dimensional, we considered slightly larger run-sizes $n = 50, 75, 100$ and $125$, however, the number of simulations and the candidates for the boundary values of $\delta$ in the likelihood optimization were kept the same as in Examples~1 and 2.

\begin{table}[!h]\centering \caption{Median and $(P_5, P_{95})$ of $\xi_{I,M}^0$ values for the proposed approach (denoted by ``$lb$'') and the popular approach (denoted by ``$mle$'') applied to the borehole model.}
{\footnotesize
\begin{tabular}{|c|c|c|c|c|}
\hline
                & $n=50$ & $n=75$ & $n=100$ & $n=125$\\
\hline
$\hat{\delta}_{mle} \ge 10^{-5}$  & 0.62 (0.21, 1.26)& 1.27 (0.69, 1.58)& 1.55 (1.11, 1.91)& 1.86 (1.44, 2.30)\\
$\hat{\delta}_{mle} \ge 10^{-10}$ & 0.48 (-1.59, 1.12)& 0.65 (-1.05, 1.56)& 0.83 (-1.47, 1.66)& 1.33 (-0.41, 2.15)\\
\hline
$\delta_{lb} (M=1)$ & -18.47 (-19.45, -16.66)&  -16.18 (-17.06, -14.27)&  -13.93 (-15.19, -12.46)&  -14.74 (-16.00, -13.47)\\
\hline
\end{tabular}\label{tab:borehole}
}
\end{table}

As expected, the interpolation error of the GP fits obtained using the popular approach slightly decreases by lowering the boundary value of the nugget parameter (from $10^{-5}$ to $10^{-10}$) in the optimization problem. The proposed method leads to predictors with significantly higher interpolation accuracy. The percentiles of $\xi_{I,M}^0$ in the last row of Table~\ref{tab:borehole} also suggest that most of the correlation matrices are well-conditioned for the $\theta$ values near $\hat{\theta}_{mle}$, and $\delta_{lb}(\hat{\theta}_{mle})=0$. That is, the iterative approach is not needed to further improve the interpolation accuracy.

Table~\ref{tab:borehole_pred} summarizes the prediction accuracy values (i.e., $\xi_{P,M}$) for 50 simulations. The test set for computing $\xi_{P,M}^0$ is a randomly chosen $8000$-point maximin Latin hypercube design. It is clear from Table~\ref{tab:borehole_pred} that more accurate prediction can be achieved by lowering the $\delta$ value in the optimization process of the popular approach, and certainly the proposed approach results in the best prediction at unsampled points among the three cases considered here.

\begin{table}[!h]\centering \caption{Median and $(P_5, P_{95})$ of $\xi_{P,M}^0$ values for the predictors in the proposed approach (denoted by ``$lb$'') and the popular approach (denoted by ``$mle$'') applied to the borehole model.}
{\footnotesize
\begin{tabular}{|c|c|c|c|c|}
\hline
                & $n=50$ & $n=75$ & $n=100$ & $n=125$\\
\hline
$\hat{\delta}_{mle} \ge 10^{-5}$  & 4.01 (3.73, 4.37)& 4.04 (3.68, 4.34)& 4.13 (3.77, 4.53)& 4.23 (3.93, 4.56)\\
$\hat{\delta}_{mle} \ge 10^{-10}$ & 3.90 (3.55, 4.32)& 3.81 (3.48, 4.16)& 3.74 (3.27, 4.04)& 3.77 (3.43, 4.07)\\
\hline
$\delta_{lb} (M=1)$               & 3.73 (3.37, 4.08)& 3.57 (3.22, 3.86)& 3.37 (2.94, 3.82)& 3.64 (3.25, 4.03)\\
\hline
\end{tabular}\label{tab:borehole_pred}
}
\end{table}
}
\end{example}

\begin{example}{\rm We now revisit the tidal power example in Section~2. The computer simulator (a version of FVCOM) is expensive and cannot be evaluated at numerous coordinates. Each of the runs presented here required approximately one hour to run on $4$ processors in parallel on the Atlantic Computational Excellence network (ACEnet) mahone cluster. While this is not particularly onerous on a large cluster, the grid resolution used in KMLH is about 200 m (length of a side in a triangle). A realistic model of 20 m sided triangular grid and with 10 vertical layers to model 3D flow would increase the computational expense by a factor of $5120$, making each individual simulator run roughly $10$ times more costly than the generation of the entire data set examined here. The ocean modelers believe that the simulator is deterministic up to the machine precision and they are interested in an emulator that interpolates the simulator.

A total of $533$ runs (on a $13 \times 41$ grid) were used to obtain the data displayed in Figure~\ref{fig:MP_coarse_grid}. We use this data to compare our results. The goal is to build an emulator of the computer model using a fraction of the budget ($533$ runs) that provides the best approximation of the simulator.
\begin{figure}[h!]\centering
  \includegraphics[height=3.35in,width=5.5in]{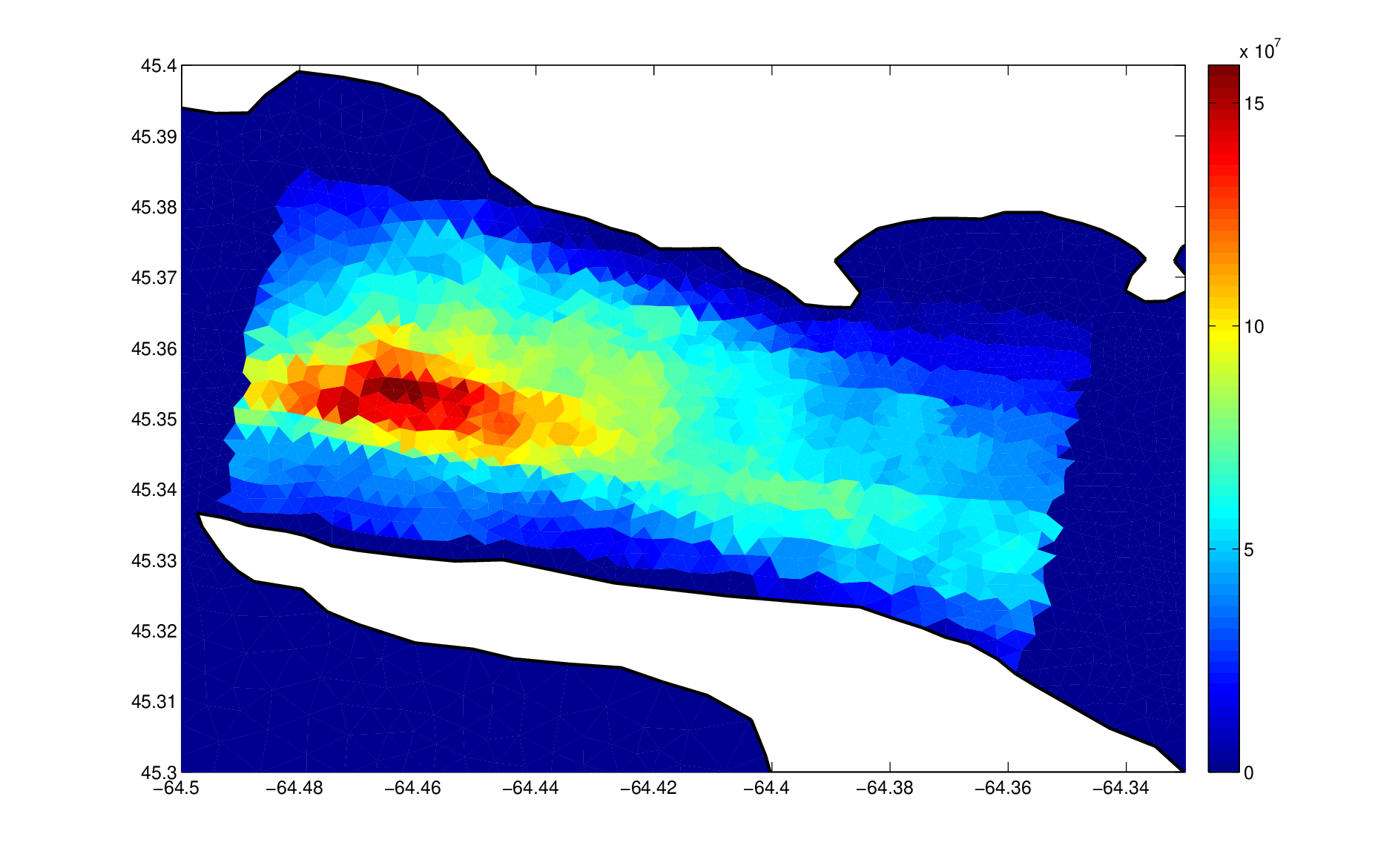} \\[-0.35cm]
\caption{FVCOM outputs (average extractable power) over a coarse grid in the Minas Passage. A colored version of the figure is available online. }\label{fig:MP_coarse_grid}
\end{figure}

We used a maximin based coverage design (Johnson, Moore and Ylvisaker 1990) to choose a subset of $n=100$ points from these $533$ points to constitute a space-filling design. The contours from both the predicted surface and the true simulator (based on the $13 \times 41$ grid) are shown in Figure~\ref{fig:motivating_eg_iter}. For successful implementation of the popular approach outlined in Sections~3.1 and 3.3, the likelihood optimization took place in the parameter space $\delta \in (10^{-5}, 1)$ and $\theta \in (0, \infty)^2$, and the parameter estimates for the GP fit are $\hat{\theta}_{mle} = (163.18, 50.66)$ and $\hat{\delta}_{mle} = 0.0462$ (see Figure~6(a)). The parameter estimates for the GP model fitted using the proposed approach with $M=1$ are $\hat{\theta}_{mle} \approx (788.54, 221.18)$ and $\delta_{lb}(\hat{\theta}_{mle}) \approx 0$ (see Figure~6(b)).

}\label{exam:exam3}\end{example}

%
\begin{figure}[h!]\centering
\subfigure[Popular fit with MLE]{\label{fig:eg2a} \includegraphics[height=2.5in,width=3.20in]{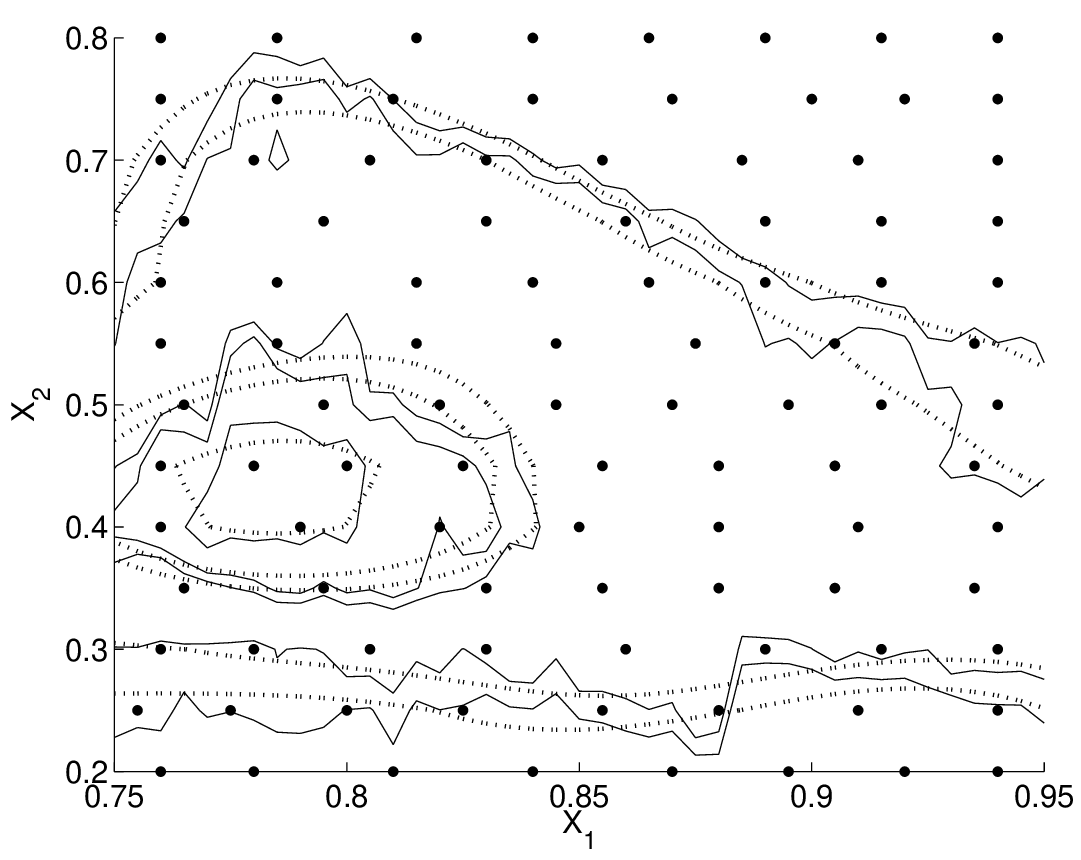}}\hspace{-1cm}
\subfigure[$M=1$ (with lower bound)]{\label{fig:eg2b} \includegraphics[height=2.5in,width=3.20in]{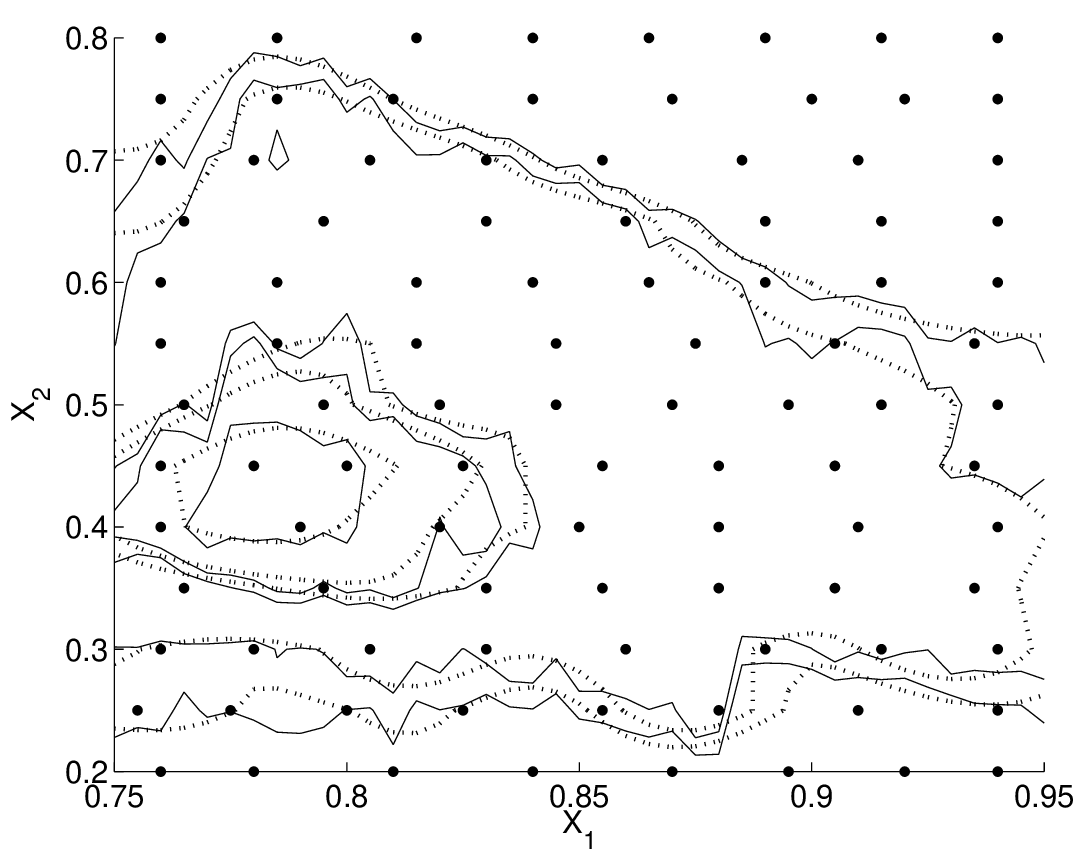}}
\caption{Dots denote the design points, the solid curves denote the contours for the true simulator based on the $13 \times 41$ grid, the dashed curves show the contours of the GP fits.}\label{fig:motivating_eg_iter}
\end{figure}

Figure~\ref{fig:motivating_eg_iter} shows that the GP based emulator obtained using the proposed approach (Figure~\ref{fig:motivating_eg_iter}(b)) is less smooth as compared to
the emulator obtained via the popular approach (Figure~\ref{fig:motivating_eg_iter}(a)). The interpolation errors for the GP fits obtained with the popular method and the proposed approach are $\xi_{I,M}^0= 6.63$ and $-26.58$ respectively. That is, the proposed approach is better at approximating the interpolator. In terms of predicting the power surface at unsampled points (i.e., the rest of $433$ points), the prediction error values for both the popular and proposed approaches are somewhat close, $\xi_{P,M}^0 \approx 10$. Moreover, when using the popular approach, the maximum predicted power obtained by evaluating $\max\{\hat{y}(x), x \in \chi\}$ is approximately $1.4 \cdot 10^8$ W with the maximizer being (0.7850, 0.4500), whereas if we use the proposed approach the maximum predicted power is approximately $1.6 \cdot 10^8$ W observed at (0.7900, 0.4500).

\section{Discussion}\label{sec:discussion}

Assuming that the underlying computer simulator is deterministic up to the machine precision and the statistician is certain about the suitability of a GP model with Gaussian correlation as the emulator, fitting the model to a data set with $n$ points in $d$-dimensional input space requires computation of the determinant and inverse of $n \times n$ correlation matrices for several $\theta$ values. In Section~\ref{sec:nugget}, we conducted a simulation study to explore space-filling designs (specifically maximin Latin hypercube designs) for different combinations of $(n, d)$ that can lead to near-singular correlation matrices. In Section~\ref{sec:iterative}, we proposed an iterative approach, that is also a generalization of the popular approach, to construct a new predictor $\hat{y}_{\delta, M}$ that has higher interpolation accuracy as compared to $\hat{y}_{\delta}$ --- the predictor from the popular approach. Lemmas~1, 2 and 3 show that $\hat{y}_{\delta, M}$ converges to the BLUP as the number of iterations ($M$) increases. The lower bound $\delta_{lb}$, proposed in Section~\ref{sec:nugget}, also allows us to use a non-zero nugget only when needed, and in
this case minimizes the number of iterations required to reach the desired interpolation accuracy.

There are a few important remarks worth noting. First, the methodology developed here can also be adapted to the Bayesian framework. For computing the posterior of the parameters and the predictor, $|R|$ and $R^{-1}$ need to be computed for several realizations of $\theta$, and a nugget is often used to overcome the near-singularity of $R$ (e.g., Taddy et al. 2009). The proposed lower bound $\delta_{lb}$ can be used for defining a prior for $\delta$, i.e., the search should be limited to $[\delta_{lb}, 1)$. One can also use the iterative approach to further improve the interpolation and/or prediction accuracy.

Second, we used the squared exponential correlation ($p_k=p=2$ for all $k$) in the GP model because of its popularity and good theoretical properties. It turns out the GP model with other power exponential correlation (i.e., $p_k=p<2$) may lead to predictors with larger MSE and sometimes worse fits as compared to that of the GP models with the Gaussian correlation. Recall that the near-singularity of $R$ occurs because (a) at least two of the design points (say $x_i$ and $x_j$) are close together in the input space, and/or (b) the hyper-parameters $\theta_k, k=1,...,d$ are very close to zero, i.e., $\sum_{k=1}^d \theta_k |x_{k,i}-x_{k,j}|^{p_k}\approx 0$. This makes a few of the rows of $R$ very similar, and will happen even if $p_k<2$. That is, the ill-conditioning problem may also occur when other power exponential correlation functions (i.e., $p_k$'s are same and less than $2$ or $p_k$'s are different and less than $2$) are used. A closer investigation reveals that with $p_k=p <2$, near-singular cases occur very frequently in the sequential design setup. However, for the space-filling designs, it is rather fascinating that the occurrence of near-singular cases is substantially reduced by even a small reduction in the power from $p = 2$ to $p = 1.99$. We suspect this is due to the limiting behaviour of the Gaussian correlation in the family of power exponential correlation functions $p \in (0, 2]$.

In conclusion, when fitting a GP model to a data set obtained from a deterministic computer model with nearly--singular correlation matrices,  we recommend using $\delta_{lb}$ - the lower bound on the nugget, along with the iterative approach with the number of iterations, $M$, chosen according to the desired interpolation accuracy.


\section*{Acknowledgments}
We would like to thank the AE and two anonymous referees for many useful comments and suggestions that lead to significant improvement of the paper. This work was partially
supported by Discovery grants from the Natural Sciences and Engineering Research Council of Canada.

\begin{center}
{\textbf{REFERENCES}}
\end{center}

\begin{description}
\item {\sc Ababou, R., Bagtzoglou, A. C. \normalfont{and} \sc Wood, E. F.} (1994). On the condition number of covariance matrices in kriging, estimation, and simulation of random fields. \emph{Mathematical Geology}, 26, 99 - 133.

\item {\sc Andre, J., Siarry, P. \normalfont{and}\sc Dognon, T.} (2000). An improvement of the standard genetic algorithm fighting premature convergence. \emph{Advances in Engineering Software}, 32, 49 - 60.

\item {\sc Aslett, R., Buck, R. J., Duvall, S. G., Sacks, J. \normalfont{and} \sc Welch, W. J.} (1998). Circuit optimization via sequential computer experiments: design of an output buffer. \emph{Applied Statistics}, 47(1), 31 -- 48.

\item {\sc Barton, R. R. \normalfont{and} \sc Salagame, R. R.} (1997). Factorial hypercube designs for spatial correlation regression. \emph{Journal of Applied Statistics}, 24, 453 - 473.

\item {\sc Bastos, L. \normalfont{and} \sc O'Hagan, A.} (2009). Diagnostics for Gaussian Process Emulators. \emph{Technometrics}, 51, 4, 425 - 438.

\item {\sc Bergmann F. T. \normalfont{and} \sc Sauro H. M.} (2008). Comparing simulation results of SBML capable simulators. \emph{Bioinformatics}, 24, 1963 -- 1965.

\item {\sc Booker, A. J., Dennis Jr., J. E., Frank, P. D., Serafini,  D. B., Torczon, V. \normalfont{and} \sc Trosset, M. W.} (1999). A rigorous framework for optimization of expensive functions by surrogates. \emph{Structural and Multidisciplinary Optimization}, 17, 1 - 13.

\item {\sc Booker, A.} (2000). Well-conditioned Kriging models for optimization of computer simulations. \emph{Mathematics and Computing Technology Phantom Works - The Boeing Company}, M\&CT-TECH-00-002.

\item {\sc Branin, F. K.} (1972). A widely convergent method for finding multiple solutions of simultaneous nonlinear equations. \emph{IBM J. Res. Develop.}, 504 - 522.

\item {\sc Chen, C., Beardsley, R. C. \normalfont{and} \sc Cowles, G.} (2006). An unstructured grid, finite-volume coastal ocean model (FVCOM) system. \emph{Oceanography}, 19, 78 - 89.

\item {\sc Golub, G. H. \normalfont{and} \sc Van Loan, C. F.} (1996). \emph{Matrix Computations}. Johns Hopkins University Press, Baltimore, MD.

\item {\sc Gramacy, R. B. \normalfont{and} \sc Lee, H. K. H.} (2008). Bayesian treed Gaussian process models with an application to computer modeling. \emph{J. Amer. Statis. Ass.}, 103(483), 1119-1130.

\item {\sc Greenberg, D.} (1979). A numerical model investigation of tidal phenomena in the Bay of Fundy and Gulf of Maine. \emph{Marine Geodesy}, 2, 161 - 187.

\item {\sc Huang, D., Allen, T.T.,  Notz, W.I. \normalfont{and} \sc Miller, R.A.} (2006). Sequential Kriging optimization using multiple fidelity evaluations, \emph{Struct. Multidisc Optim.}, 32, 369 - 382.

\item {\sc Johnson, M. E., Moore, L. M., \normalfont{and} \sc Ylvisaker, D.} (1990). Minimax and maximin distance designs. \emph{Journal of Statistical Planning and Inference}, 26, 131 - 148.

\item {\sc Jones, D., Schonlau, M., \normalfont{and} \sc Welch, W.} (1998).  Efficient Global Optimization of Expensive Black-Box Functions.  \emph{Journal of Global Optimization}, 13, 455 - 492.

\item {\sc Joseph, V. R., Hung, Y., \normalfont{and} \sc A. Sudjianto.} (2008). Blind Kriging: A New Method for Developing Metamodels. \emph{Journal of Mechanical Design}, 130(3), 31 -- 102.

\item {\sc Lebedev, V. I.} (1997). \emph{An introduction to functional analysis in computational mathematics}, Birkh\"auser, Boston.

\item {\sc Karsten, R., McMillan, J., Lickley, M. \normalfont{and} \sc Haynes, R.} (2008). Assessment of tidal current energy for the Minas Passage, Bay of Fundy. \emph{Proceedings of the Institution of Mechanical Engineers, Part A: Journal of Power and Energy}, 222, 493 - 507.

\item {\sc Kumar, B. \normalfont{and} \sc Davidson, E. S.} (1978). Performance evaluation of highly concurrent computers by deterministic simulation, \emph{Communications of the ACM}, 21 (11), 904 -- 913.

\item {\sc Medina, J.S., Moreno, M.G. \normalfont{and} \sc Royo, E.R.} (2005). Stochastic vs deterministic traffic simulator. Comparative study for its use within a traffic light cycles optimization architecture, \emph{In proc. IWINAC} (2), 622 -- 631.

\item {\sc Neal, R. M.} (1997). Monte Carlo implementation of Gaussian process models for Bayesian regression and classification. Tech Rep. 9702, \emph{Dept. of Statistics, Univ. of Toronto, Canada}.

\item {\sc Neumaier, A.} (1998). Solving ill-conditioned and singular linear systems: A tutorial on regularization. \emph{SIAM Review}, 40, 636 - 666.

\item {\sc Oakley, J.} (2004). Estimating percentiles of computer code outputs. \emph{Appl. Statist.}, 53, 83 - 93.

\item {\sc Poole, D. \normalfont{and} \sc A. E. Raftery.} (2000). Inference for deterministic simulation models: the Bayesian melding approach. \emph{J. Amer. Statis. Ass.}, 95 (452), 1244-1255.

\item{\sc Ranjan, P., Bingham, D. \normalfont{and} \sc Michailidis, G.} (2008). Sequential experiment design for contour estimation from complex computer codes. \emph{Technometrics}, 50, 527 - 541.

\item{\sc Rasmussen, C. E.} and {\sc Williams, C. K. I.} (2006). \emph{Gaussian Processes for Machine Learning}. The MIT Press.

\item {\sc Sacks, J., Welch, W. J., Mitchell, T. J. \normalfont{and} \sc Wynn, H. P.} (1989). Design and analysis of computer experiments. \emph{Statistical Science}, 4, 409 - 423.

\item {\sc Santner, T. J., Williams, B. J. \normalfont{and} \sc Notz, W.} (2003) \emph{The Design and Analysis of Computer Experiments}. Springer-Verlag Inc., New York, NY.

\item {\sc Schonlau, M., Welch, W. \normalfont{and} \sc Jones, D.} (1998). Global versus local search in constrained optimization of computer models. \emph{New Developments and Applications in Experimental Design, Institute of Mathematical Statistics Lecture Notes, Hayward, California}, 34, 11 - 25

\item {\sc Stein, M.} (1987). Large Sample Properties of Simulations Using Latin Hypercube Sampling. \emph{Technometrics}, 29, 143 -- 151.

\item {\sc Stein, M. L.} (1999). \emph{Interpolation of Spatial Data: Some Theory for Kriging}. Springer, NY.

\item {\sc Su, H, Nelder, JA, Wolbert, P \normalfont{and} \sc Spence, R} (1996). Application of generalized linear models to design improvement of an engineering artefact. \emph{Qual Reliab Engng Int}, 12, 101-112.

\item {\sc Taddy, M. A., Lee, H. K. H., Gray, G. A. \normalfont{and} \sc Griffin, J.} (2009). Bayesian guidance for robust pattern search optimization. \emph{Technometrics} (to appear).

\item {\sc Tikhonov, A. N.} (1993). Solution of incorrectly formulated problems and the regularization method. \emph{Soviet Math. Doklady}, 4, 1035 - 1038.

\item {\sc Van Beers, W. C. M., \normalfont{and} \sc Kleijnen, J. P.C.} (2004). Kriging interpolation in simulation: a survey. \emph{In Proceedings of the 2004 Winter Simulation Conference}, ed. R.G. Ingalls, M.D. Rossetti, J.S. Smith, and B.A. Peter, 113 - 121. Piscataway, New Jersey: Institute of Electrical and Electronics Engineers.

\item {\sc Yang, X. -S.} (2010). Test problems in optimization, \emph{Engineering Optimization: An Introduction with Metaheuristic Applications (Eds Xin-She Yang), John Wiley \& Sons}.

\end{description}

\end{document}